\documentclass[11pt,epsf]{article}
 \usepackage{amsmath}
 \usepackage{amssymb}
 \usepackage{graphicx}
 \usepackage[merge,numbers,compress]{natbib}
 \usepackage[T1]{fontenc}
 \usepackage{booktabs}
 \usepackage{xcolor} 
 \usepackage{xspace}
 \usepackage{dcolumn}
 \usepackage{placeins}
 \usepackage[colorlinks=true,citecolor=blue!50!black,linkcolor=black]{hyperref}
 \usepackage{caption}
 \usepackage
 [subrefformat=parens,position=top,skip=-15pt,margin=15pt,justification=justified,singlelinecheck=false]
 {subcaption}
 \usepackage{multirow}
 \usepackage{comment}

\newcommand{\process}{\Pp\Pp\to\mu^+\nu_\mu\Pe^+\nu_{\Pe}\Pj\Pj}
\newcommand{\processfs}{\mu^+\nu_\mu\Pe^+\nu_{\Pe}\Pj\Pj}

\newcommand{\fb}{{\ensuremath\unskip\,\text{fb}}\xspace}
\def\refeq#1{\mbox{(\ref{#1})}}
\def\reffi#1{\mbox{Figure~\ref{#1}}}
\def\reffis#1{\mbox{Figures~\ref{#1}}}
\def\refta#1{\mbox{Table~\ref{#1}}}
\def\reftas#1{\mbox{Tables~\ref{#1}}}
\def\refse#1{\mbox{Sec.~\ref{#1}}}

\def\citere#1{\mbox{Ref.~\cite{#1}}}
\def\citeres#1{\mbox{Refs.~\cite{#1}}}

\newcommand{\ie}{\emph{i.e.}\ }
\newcommand{\eg}{\emph{e.g.}\ }
\newcommand{\cf}{c.f.\ }

\let\origfootnote\footnote
\renewcommand{\footnote}[1]{\kern.06em\origfootnote{#1}}
\newcommand{\punctfootnote}[1]{\kern-.06em\origfootnote{#1}}

\def\be{\begin{equation}}
\def\ee{\end{equation}}

\newcommand{\PH}{\ensuremath{\text{H}}\xspace}
\newcommand{\Pj}{\ensuremath{\text{j}}\xspace}
\newcommand{\Pp}{\ensuremath{\text{p}}\xspace}
\newcommand{\Pe}{\ensuremath{\text{e}}\xspace}
\newcommand{\Pb}{\ensuremath{\text{b}}\xspace}

\newcommand{\Pt}{\ensuremath{\text{t}}\xspace}
\newcommand{\Pu}{\ensuremath{\text{u}}\xspace}
\newcommand{\Pd}{\ensuremath{\text{d}}\xspace}
\newcommand{\Ps}{\ensuremath{\text{s}}\xspace}
\newcommand{\Pc}{\ensuremath{\text{c}}\xspace}

\newcommand{\PW}{\ensuremath{\text{W}}\xspace}
\newcommand{\PZ}{\ensuremath{\text{Z}}\xspace}

\newcommand{\Mt}{\ensuremath{m_\Pt}\xspace}
\newcommand{\MH}{\ensuremath{M_\PH}\xspace}
\newcommand{\MWOS}{\ensuremath{M_\PW^\text{OS}}\xspace}
\newcommand{\MW}{\ensuremath{M_\PW}\xspace}
\newcommand{\MZOS}{\ensuremath{M_\PZ^\text{OS}}\xspace}
\newcommand{\MZ}{\ensuremath{M_\PZ}\xspace}
\newcommand{\Mb}{\ensuremath{m_\Pb}\xspace}

\newcommand{\GH}{\ensuremath{\Gamma_\PH}\xspace}

\newcommand{\GZOS}{\ensuremath{\Gamma_\PZ^\text{OS}}\xspace}

\newcommand{\GWOS}{\ensuremath{\Gamma_\PW^\text{OS}}\xspace}

\newcommand{\MeV}{\ensuremath{\,\text{MeV}}\xspace}
\newcommand{\GeV}{\ensuremath{\,\text{GeV}}\xspace}
\newcommand{\TeV}{\ensuremath{\,\text{TeV}}\xspace}

\newcommand{\alphas}{\ensuremath{\alpha_\text{s}}\xspace}
\newcommand{\gs}{\ensuremath{g_\text{s}}\xspace}
\newcommand{\order}[1]{\ensuremath{\mathcal{O}{\left(#1\right)}}\xspace}

\newcommand{\GF}{\ensuremath{G_\mu}}

\newcommand{\ptsub}[1]{\ensuremath{p_{\text{T},#1}}\xspace}

\newcommand{\MVOS}{\ensuremath{M_{\text{V}}^\text{OS}}\xspace}%
\newcommand{\GVOS}{\ensuremath{\Gamma_{\text{V}}^\text{OS}}\xspace}%

\newcommand{\newc}{\newcommand}
\newc{\bi}{\begin{itemize}}
\newc{\ei}{\end{itemize}}
\newc{\benu}{\begin{enumerate}}
\newc{\eenu}{\end{enumerate}}
\newc{\bc}{\begin{center}}
\newc{\ec}{\end{center}}
\newc{\bfig}{\begin{figure}}
\newc{\efig}{\end{figure}}
\newc{\qbar}{\bar{q}}
\newc{\go}{\tilde{g}}
\newc{\PB}{\textsc{Powheg-Box}}

\newcommand{\Recola}{{\sc Recola}\xspace}
\newcommand{\Sherpa}{{\sc Sherpa}\xspace}

\newcommand{\MoCaNLORecola}{{\sc  MoCaNLO{+}Recola}\xspace}
\newcommand{\Rivet}{{\sc Rivet}\xspace}
\newcommand{\Amegic}{A\protect\scalebox{0.8}{MEGIC}\xspace}
\newcommand{\Comix}{C\protect\scalebox{0.8}{OMIX}\xspace}

\newcommand{\mocanlo}{{\sc MoCaNLO}\xspace}
\newcommand{\collier}{{\sc Collier}\xspace}

\newcommand{\rT}{{\mathrm{T}}}
\newcolumntype{.}{D{.}{.}{-1}}
\newcolumntype{d}[1]{D{.}{.}{#1}}

\newcommand{\EW}{\ensuremath{\text{EW}}}

\newcommand{\EWvirt}{\ensuremath{\text{EW}_\text{virt}}}

\newcommand{\MCatNLO}{\text{\textsc{Mc@Nlo}}\xspace}

\newcommand{\mr}[1]{\ensuremath{\mathrm{#1}}}

\newcommand{\muR}{\ensuremath{\mu_{\mr{R}}}}
\newcommand{\muF}{\ensuremath{\mu_{\mr{F}}}}

\colorlet{tableoverheadcolor}{gray!37.5}
\colorlet{tableheadcolor}{gray!25}
\colorlet{tablerowcolor}{gray!12.5}

\newcommand{\lsim}
{\;\raisebox{-.3em}{$\stackrel{\displaystyle <}{\sim}$}\;}
\newcommand{\gsim}
{\;\raisebox{-.3em}{$\stackrel{\displaystyle >}{\sim}$}\;}

\newlength{\width}
\newlength{\height}

\def\draftdate{\relax}
\def\mda{\relax}
\def\mua{\relax}
\def\mla{\relax}
\def\draft{
\def\thtystars{******************************}
\def\sixtystars{\thtystars\thtystars}
\typeout{}
\typeout{\sixtystars**}
\typeout{* Draft mode!
         For final version remove \protect\draft\space in source file *}
\typeout{\sixtystars**}
\typeout{}
\def\draftdate{\today}
\def\mua{\marginpar[\boldmath\hfil$\uparrow$]%
                   {\boldmath$\uparrow$\hfil}\color{black}%
                    \typeout{marginpar: $\uparrow$}\ignorespaces}
\def\mda{\color{red}\marginpar[\boldmath\hfil$\downarrow$]%
                   {\boldmath$\downarrow$\hfil}%
                    \typeout{marginpar: $\downarrow$}\ignorespaces}
\def\mla{\marginpar[\boldmath\hfil$\rightarrow$]%
                   {\boldmath$\leftarrow $\hfil}%
                    \typeout{marginpar: $\leftrightarrow$}\ignorespaces}
\def\Mua{\marginpar[\boldmath\hfil$\Uparrow$]%
                   {\boldmath$\Uparrow$\hfil}\color{black}%
                    \typeout{marginpar: $\uparrow$}\ignorespaces}
\def\Mda{\color{red}\marginpar[\boldmath\hfil$\Downarrow$]%
                   {\boldmath$\Downarrow$\hfil}%
                    \typeout{marginpar: $\downarrow$}\ignorespaces}
\def\Mla{\marginpar[\boldmath\hfil\textcolor{red}{$\Rightarrow$}]%
                   {\boldmath\textcolor{red}{$\Leftarrow $}\hfil}%
                    \typeout{marginpar: $\leftrightarrow$}\ignorespaces}
\overfullrule 5pt
\oddsidemargin 15mm
\marginparwidth 29mm
}

\newcommand{\hl}{\vphantom{$\int_A^B$}}

\newcolumntype{C}{>{\centering\arraybackslash}p{0.105\textwidth}}
\newcolumntype{D}{>{\centering\arraybackslash}p{0.12\textwidth}}
\newcolumntype{E}{>{\centering\arraybackslash}p{0.16\textwidth}}

\begin{document}

\hypersetup{pageanchor=false}

\title{\hfill ~\\[-30mm]
\phantom{h} \hfill\mbox{\small FR-PHENO-2024-02, IPPP/24/31, MCNET-24-11}
\\[1cm]
\vspace{13mm}   \textbf{Tri-boson and WH production in the $\PW^+\PW^+\Pj\Pj$ channel: \\ predictions at full NLO accuracy and beyond}}

\date{}
\author{
Ansgar Denner$^{1\,}$\footnote{E-mail:
  \texttt{ansgar.denner@uni-wuerzburg.de}},
Mathieu Pellen$^{2\,}$\footnote{E-mail:
  \texttt{mathieu.pellen@physik.uni-freiburg.de}},
Marek Sch\"onherr$^{3\,}$\footnote{E-mail:
  \texttt{marek.schoenherr@durham.ac.uk}},
Steffen Schumann$^{4\,}$\footnote{E-mail: \texttt{steffen.schumann@phys.uni-goettingen.de}}
\\[9mm]
{\small\it $^1$ Universit\"at W\"urzburg, Institut f\"ur Theoretische Physik und Astrophysik,} \\
{\small\it Emil-Hilb-Weg 22, \linebreak
        97074 W\"urzburg,
        Germany}\\[3mm]
{\small\it $^2$ Universit\"at Freiburg, Physikalisches Institut,} \\
{\small\it Hermann-Herder-Str. 3, 79104 Freiburg, Germany}\\[3mm]
{\small\it $^3$ Institute for Particle Physics Phenomenology, Durham University,} \\
{\small\it Durham DH1 3LE, United Kingdom}\\[3mm]
{\small\it $^4$ Georg-August-Universit\"at G\"ottingen, Institut f\"ur Theoretische Physik,} \\ %
{\small\it Friedrich-Hund-Platz 1, \linebreak
        37077 G\"ottingen,
        Germany}\\[3mm]
}

\maketitle

\begin{abstract}
\noindent
In this work, we present the first full NLO predictions for the process $\Pp\Pp\to\mu^+\nu_\mu\Pe^+\nu_\Pe\Pj\Pj$
at the LHC in a typical tri-boson phase space. The NLO corrections
reach $50\%$ at the level of the fiducial
cross section and have a very different hierarchy with respect to vector-boson-scattering phase spaces.
By comparing the cross section of the full off-shell process with the
sum of contributing on-shell electroweak-boson production subchannels, 
we find that the process is dominated by $\PW\PW\PW$ and $\PW\PH$ production, while vector-boson-scattering topologies still play a non-negligible role.
In addition, NLO QCD predictions matched to parton shower which are
supplemented by approximate electroweak corrections are provided.
For the fiducial cross section, the electroweak corrections turn out
to be small but the QCD corrections reach $47\%$. For the inclusive
cross section, matching to parton shower affects the predictions by $7\%$.
However, for differential distributions corrections due to the parton shower
can be much more sizeable, depending on the region of phase space. 

\end{abstract}
\thispagestyle{empty}
\vfill
\newpage
\setcounter{page}{1}
\hypersetup{pageanchor=true}

\tableofcontents

\section{Introduction}

The physics programme at the Large Hadron Collider (LHC) relies on the comparison of experimental measurements and theoretical predictions.
While this task might sound trivial, it is actually extremely complex
as typically different objects/concepts are considered on both sides.
In particular, the definition of the process measured or computed is at the core of these comparisons.

Experimentally, only the final states retained
and the event selection applied to them define the experimental signature and the associated process measured.
On the other hand, theoretical predictions require knowledge of
all the external states, including the initial ones, as well as the order in perturbation theory to be considered.
None of these definitions actually refer to the content of the
intermediate virtual particles as these are not physically manifest since
only their decay products can be observed experimentally.

Nonetheless, processes are usually claimed to be measured or computed based on their content of intermediate particles, \eg top-pair production or di-boson production.
This becomes meaningful when selecting phase-space regions that enhance the contributions of interest.
In addition, on the experimental side, irreducible backgrounds are often removed using theoretical inputs.
On the theory side, instead of computing the full off-shell process, several methods can be used in order to single out certain contributions.
It should be kept in mind that all these treatments rely on approximations that eventually aim at simplifying the interpretation while blurring the physical meaning of the comparisons.

For our purpose, the final state under consideration is $\mu^+\nu_\mu\Pe^+\nu_{\Pe}\Pj\Pj$, mostly known as the signature of vector-boson scattering (VBS) of same-sign W bosons, which is the golden channel for VBS measurements at the LHC~\cite{Covarelli:2021gyz}.
In such analyses, the phase-space requirements are rather extreme with
large invariant mass and rapidity separation for the two tagging jets in order to single out the electroweak (EW) production from the QCD-induced background.
On the other hand, for different phase spaces such as the one used here,
which is a simplified version of a tri-boson measurement, the process is actually dominated by Higgs-strahlung (WH) and tri-boson contributions.

On the experimental side, all three processes (VBS, tri-boson, WH) have been measured within various phase spaces and/or final states.
The production of a W boson in association with a Higgs boson has long been measured by both ATLAS and CMS~\cite{CMS:2018vqh,ATLAS:2019vrd,ATLAS:2019yhn}.
Nonetheless, until now and to the best of our knowledge, no experimental measurement has
been designed to probe WH production in the $\PW^+\PW^+\Pj\Pj$ channel.
For VBS, the same-sign WW channel has provided the first VBS measurements performed at the LHC~\cite{Aad:2014zda,Khachatryan:2014sta,ATLAS:2016snd,ATLAS:2023sua}.
For what concerns tri-boson measurements, the process has been searched for and was ultimately observed in various channels \cite{ATLAS:2016jeu,CMS:2020hjs,ATLAS:2022xnu}.
However, only the latest measurement of the ATLAS collaboration~\cite{ATLAS:2022xnu} is actually sensitive to the semi-leptonic decay channel under consideration here.
Also, it is worth mentioning that in the latter analysis, a tension has been found between the experimental data and the Standard Model (SM) expectations.

On the theory side, the process
$\Pp\Pp\to\mu^+\nu_\mu\Pe^+\nu_{\Pe}\Pj\Pj$ is known at full
next-to-leading-order (NLO) accuracy~\cite{Biedermann:2017bss,Dittmaier:2023nac} in a VBS phase
space.\punctfootnote{NLO QCD predictions in an inclusive phase space based
  on the same calculation have been presented in
  \citere{Ballestrero:2018anz}. Previous calculations~\cite{Jager:2009xx,Jager:2011ms,Denner:2012dz} relied on the VBS approximation and are thus not valid in different phase spaces.}  The EW component has been found to
feature particularly large EW
corrections~\cite{Biedermann:2016yds},\punctfootnote{We note that these EW
  corrections are publicly available \cite{Chiesa:2019ulk} in the resonance-aware version of {\sc Powheg}~\cite{Jezo:2015aia}.
  While there is in principle no restriction on
  the phase space in this implementation, it has only been tested in VBS phase space. Other regions might suffer from inefficiencies.} while they turned out
to be of the expected size for off-shell tri-boson production in the
leptonic channels~\cite{Schonherr:2018jva,Dittmaier:2019twg}.
For these processes, the state of the art is NLO
QCD+EW in a fully off-shell calculation \cite{Dittmaier:2019twg}.  
For WH production with on-shell Higgs and leptonically decaying
W~boson, the inclusive cross section has been computed in the
threshold limit at ${\rm N^3LO}$ in QCD~\cite{Kumar:2014uwa}, while it is available fully differentially~\cite{Ferrera:2011bk,Ferrera:2013yga} at NNLO QCD accuracy.
The latter calculation has been matched to a parton shower~\cite{Astill:2016hpa}.
Furthermore, soft-gluon resummation results are also available for
this process~\cite{Dawson:2012gs,Alioli:2019qzz}. Concerning EW corrections, full NLO results for on-shell Higgs
production exist \cite{Denner:2011id}  and have been combined with a
QCD+QED parton shower~\cite{Granata:2017iod} in the
{\sc Powheg} framework~\cite{Nason:2004rx,Frixione:2007vw,Alioli:2010xd}.
To the best of our knowledge, there exist no specific computations in the literature discussing off-shell Higgs-strahlung processes.
In \citere{Hoeche:2014rya}, triple-boson production through
Higgs~strahlung was studied in an on-shell approximation with NLO multijet merged predictions.
Finally, the QCD corrections to the QCD background are known for some
time \cite{Melia:2010bm,Campanario:2013gea}, and an implementation at
NLO QCD plus parton shower accuracy~\cite{Melia:2011gk} is available in {\sc
  Powheg}~\cite{Nason:2004rx,Frixione:2007vw,Alioli:2010xd}.

The process $\Pp\Pp\to\mu^+\nu_\mu\Pe^+\nu_{\Pe}\Pj\Pj$ is interesting for several reasons.
First and as argued before, the process shares several subprocesses with different physics aspects. 
Second, WH and tri-boson production have never been computed for this final state at full NLO accuracy.
In that respect, it is particularly interesting to reconcile several findings on the size of various corrections for different processes in the literature.
In view of recent experimental analyses~\cite{ATLAS:2022xnu}, an
improvement and reassessment of the SM predictions is in order.

To that end, we have computed the full NLO corrections to
$\Pp\Pp\to\mu^+\nu_\mu\Pe^+\nu_{\Pe}\Pj\Pj$ in a triboson phase space
which turned out to enhance both triboson and WH contributions. 
We remark that full NLO accuracy for
processes involving more than two coupling-constant orders has only
been obtained for few cases in the on-shell description
~\cite{Frixione:2015zaa,Frederix:2017wme}, in the narrow-width
approximation~\cite{Stremmer:2024ecl} and for full off-shell
processes~\cite{Biedermann:2017bss,Frederix:2016ost,Reyer:2019obz,
Denner:2021hsa,Denner:2021hqi,Lindert:2022ejn,Denner:2023eti,Dittmaier:2023nac}.
To complement our fixed-order results and to account for multiple QCD
emissions, we provide predictions compiled with
\Sherpa~\cite{Bothmann:2019yzt,Gleisberg:2008ta} at NLO QCD matched
with the parton shower thereby including approximate NLO EW
corrections. The latter are treated in the so-called EW virtual
approximation, first presented in \citere{Kallweit:2015dum} and
applied to a variety of processes in the
meantime~\cite{Kallweit:2017khh,Brauer:2020kfv,Bothmann:2021led}, that
captures the exact NLO EW virtual corrections and integrated
approximate real-emission subtractions but discards hard
real-emission configurations. In turn, we here present the
current state of the art in fixed-order and parton-shower-evolved predictions
for $\Pp\Pp\to\mu^+\nu_\mu\Pe^+\nu_{\Pe}\Pj\Pj$ production at the LHC.

The article is organised as follows:
in \refse{sec:features}, all definitions and details of the calculations are provided.
Section \ref{sec:results} is devoted to the discussion of numerical results.
These range from leading-order (LO) studies to full NLO predictions, including detailed analyses
on the impact of off-shell contributions, to NLO QCD parton-shower matched results.
Finally, \refse{sec:conclusion} contains a summary and concluding remarks.

\section{Features of the calculations}
\label{sec:features}

\subsection{LO contributions}

\begin{figure}
\centering
\begin{subfigure}{0.3\textwidth}
\captionsetup{skip=0pt}
\caption{}
\centering
\includegraphics[page=1,width=1.\linewidth]{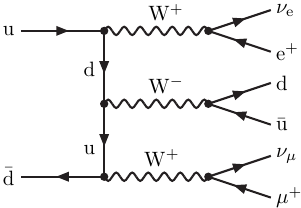}
\label{diag:LO:WWW}
\end{subfigure}
\hspace*{0.03\textwidth}
\begin{subfigure}{0.3\textwidth}
\captionsetup{skip=0pt}
\caption{}
\centering
\includegraphics[page=2,width=1.\linewidth]{diagrams/fdiagrams.pdf}
\label{diag:LO:Higgs}
\end{subfigure}
\hspace*{0.03\textwidth}
\begin{subfigure}{0.3\textwidth}
\captionsetup{skip=0pt}
\caption{}
\centering
\includegraphics[page=7,width=1.\linewidth]{diagrams/fdiagrams.pdf}
\label{diag:LO:WZ}
\end{subfigure}
\par
\begin{subfigure}{0.3\textwidth}
\captionsetup{skip=0pt}
\caption{}
\centering
\includegraphics[page=3,width=1.\linewidth]{diagrams/fdiagrams.pdf}
\label{diag:LO:quartic}
\end{subfigure}
\hspace*{0.06\textwidth}
\begin{subfigure}{0.3\textwidth}
\captionsetup{skip=0pt}
\caption{}
\centering
\includegraphics[page=6,width=1.\linewidth]{diagrams/fdiagrams.pdf}
\label{diag:LO:QCDs}
\end{subfigure}
\par
\begin{subfigure}{0.3\textwidth}
\captionsetup{skip=0pt}
\caption{}
\centering
\includegraphics[page=4,width=1.\linewidth]{diagrams/fdiagrams.pdf}
\label{diag:LO:VBS}
\end{subfigure}
\hspace*{0.06\textwidth}
\begin{subfigure}{0.3\textwidth}
\captionsetup{skip=0pt}
\caption{}
\centering
\includegraphics[page=5,width=1.\linewidth]{diagrams/fdiagrams.pdf}
\label{diag:LO:QCDt}
\end{subfigure}
\par
\begin{subfigure}{0.3\textwidth}
\captionsetup{skip=0pt}
\caption{}
\centering
\includegraphics[page=8,width=1.\linewidth]{diagrams/fdiagrams.pdf}
\label{diag:LO:VBSuu}
\end{subfigure}
\hspace*{0.06\textwidth}
\begin{subfigure}{0.3\textwidth}
\captionsetup{skip=0pt}
\caption{}
\centering
\includegraphics[page=9,width=1.\linewidth]{diagrams/fdiagrams.pdf}
\label{diag:LO:QCDtuu}
\end{subfigure}
        \caption{Examples of tree-level diagrams contributing to
          $\Pp\Pp\to\mu^+\nu_\mu\Pe^+\nu_{\Pe}\Pj\Pj$ at
          $\order{e^6}$ (a, b, c, d, f, h)
          and $\order{\gs^2e^4}$ (e, g, i).}
\label{diag:LO}
\end{figure}

The process under investigation in this work is
\begin{align}
\label{eq:process}
 \Pp\Pp\to\mu^+\nu_\mu\Pe^+\nu_{\Pe}\Pj\Pj
\end{align}
at the LHC.  This process is of particular interest as it contains
contributions with three resonant W~bosons,
$\Pp\Pp\to\PW^+(\to\mu^+\nu_\mu)\PW^+(\to\Pe^+\nu_{\Pe})\PW^-(\to\Pj\Pj)$,
as shown in \reffi{diag:LO:WWW}.  
The process also involves Higgs strahlung, \ie
$\Pp\Pp\to\PW^+(\to\mu^+\nu_\mu)\PH[\to\PW^+(\to\Pe^+\nu_{\Pe})\PW^-(\to\Pj\Pj)]$
(as well as  $\Pe^+\leftrightarrow\mu^+$ and $\nu_\Pe\leftrightarrow\nu_\mu$),
and $\PW^+\PZ$ production as subprocesses  (for representative Feynman
diagrams see \reffis{diag:LO:Higgs} and \ref{diag:LO:WZ}).
In the phase space considered in this article, the triple-resonant WWW contribution
dominates with roughly $50\%$, the Higgs-strahlung process
contributes about $40\%$,
while the $\PW^+\PZ$ contribution is very small.
Besides, many other types of
contributions are present in the EW process at order
$\order{\alpha^6}$, such as diagrams with quartic gauge
couplings (\reffi{diag:LO:quartic}) or VBS ones
(\reffis{diag:LO:VBS} and \ref{diag:LO:VBSuu}).  We note that
triple-$\PW$ production and $\PW\PH$ production only appear in partonic
channels that contain $s$-channel contributions, \ie only in quark--anti-quark
annihilation channels.

In addition, the process \refeq{eq:process} receives contributions of
order $\order{\alphas^2\alpha^4}$, which are usually
referred to as irreducible QCD background.  Examples of corresponding
diagrams are depicted in \reffis{diag:LO:QCDs}, \ref{diag:LO:QCDt},
and \ref{diag:LO:QCDtuu}.
It is worth emphasising that the relative size of the QCD background depends strongly
on the considered phase space.

Finally, interference contributions  between the EW
and QCD amplitudes arise at LO, which are of order $\order{\alphas\alpha^{5}}$.
Owing to colour algebra, these interferences are non-zero only for
those partonic channels that receive contributions from different
kinematic channels, either from $t$ and $u$~channels or from $t$ and $s$~channels.
The complete set of partonic contributions is listed in Table~1 of
\citere{Biedermann:2017bss} along with their potential interferences
at $\order{\alphas\alpha^{5}}$ and their kinematic channels.

In quark--quark and anti-quark--anti-quark channels, only $t$- or
$u$-channel diagrams like those in \reffis{diag:LO:VBSuu} and
\ref{diag:LO:QCDtuu} contribute as well as interferences between EW and
QCD amplitudes with identical final-state quarks. 

\subsection{Full NLO predictions}

\begin{figure}[t]
\begin{center}
          \includegraphics[width=0.8\linewidth]{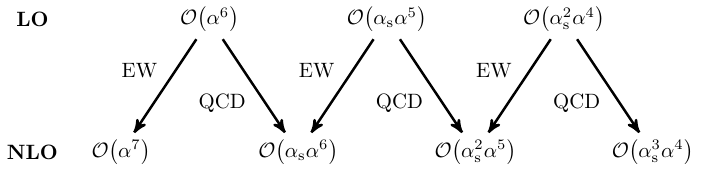}
\end{center}
        \caption{Perturbative orders contributing at LO and NLO for $\Pp\Pp\to\mu^+\nu_\mu\Pe^+\nu_{\Pe}\Pj\Pj$.}
\label{fig:allorders}
\end{figure}

\begin{figure}
\centering
\begin{subfigure}{0.3\textwidth}
\captionsetup{skip=0pt}
\caption{}
\centering
\includegraphics[page=10,width=1.\linewidth]{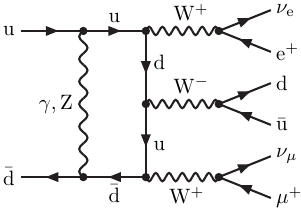}
\label{diag:loop:gs0}
\end{subfigure}
\hspace*{0.03\textwidth}
\begin{subfigure}{0.3\textwidth}
\captionsetup{skip=0pt}
\caption{}
\centering
\includegraphics[page=12,width=1.\linewidth]{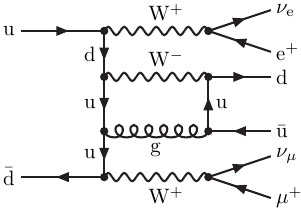}
\label{diag:loopw:gs2}
\end{subfigure}
\par
\begin{subfigure}{0.3\textwidth}
\captionsetup{skip=0pt}
\caption{}
\centering
\includegraphics[page=11,width=1.\linewidth]{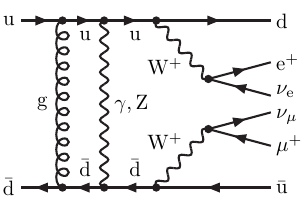}
\label{diag:loopz:gs2}
\end{subfigure}
\hspace*{0.03\textwidth}
\begin{subfigure}{0.3\textwidth}
\captionsetup{skip=0pt}
\caption{}
\centering
\includegraphics[page=13,width=1.\linewidth]{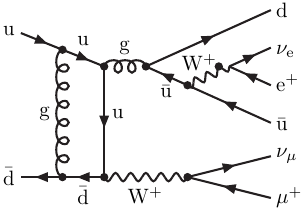}
\label{diag:loop:gs4}
\end{subfigure}
        \caption{Examples of loop diagrams contributing to
          $\Pp\Pp\to\mu^+\nu_\mu\Pe^+\nu_{\Pe}\Pj\Pj$ at
          $\order{e^8}$ (a), $\order{\gs^2e^6}$ (b,c),
          and $\order{\gs^4e^4}$ (d).}

\label{diag:loop}
\end{figure}

The three different LO contributions give rise to four terms 
contributing at NLO accuracy arising from both QCD and EW corrections.
A pictorial representation of this is provided in
\reffi{fig:allorders}.
Each NLO order is either made of pure QCD or EW corrections or a mixture of the two as for the orders $\order{\alphas\alpha^6}$ and $\order{\alphas^2\alpha^5}$.
For example, the NLO corrections of $\order{\alphas\alpha^6}$
are QCD corrections to the LO process of $\order{\alpha^6}$
and EW corrections to the LO process of $\order{\alphas\alpha^5}$ simultaneously.
In \reffi{diag:loop} we show examples of loop diagrams of the orders
$\order{e^8}$, $\order{\gs^2e^6}$, and $\order{\gs^4e^4}$
in \reffis{diag:loop:gs0},
\ref{diag:loopw:gs2} as well as \ref{diag:loopz:gs2}, and
\ref{diag:loop:gs4}, respectively.  The diagram in
\reffi{diag:loopz:gs2}, for example, can be viewed as an EW correction to the LO
diagram in \reffi{diag:LO:QCDt} or as a QCD correction to a similar LO
diagram with the gluon replaced by a photon or Z~boson. A similar
statement holds for the diagram in \reffi{diag:loopw:gs2}. Interfering
the diagrams in \reffis{diag:loopz:gs2} and \ref{diag:loopw:gs2}
with those in \reffis{diag:LO:WWW} and \ref{diag:LO:VBS} yields
contributions of order \order{\alphas\alpha^6}, while interfering them
with those in \reffis{diag:LO:QCDt} and \ref{diag:LO:QCDs} gives
contributions of order \order{\alphas^2\alpha^5}.

\begin{figure}
\centering
\begin{subfigure}{0.3\textwidth}
\captionsetup{skip=0pt}
\caption{}
\centering
\includegraphics[page=14,width=1.\linewidth]{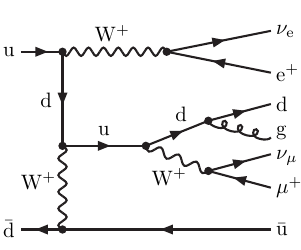}
\label{diag:real_gluon_from_quark_VV}
\end{subfigure}
\hspace*{0.03\textwidth}
\begin{subfigure}{0.3\textwidth}
\captionsetup{skip=0pt}
\caption{}
\centering
\includegraphics[page=15,width=1.\linewidth]{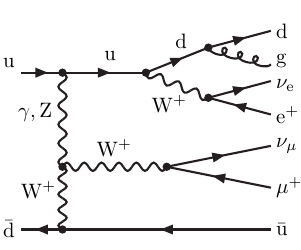}
\label{diag:real_gluon_from_quark_VV2}
\end{subfigure}
\hspace*{0.03\textwidth}
\begin{subfigure}{0.3\textwidth}
\captionsetup{skip=0pt}
\caption{}
\centering
\includegraphics[page=16,width=1.\linewidth]{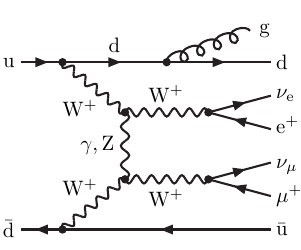}
\label{diag:real_gluon_from_quark_VBS}
\end{subfigure}
        \caption{Examples of real-radiation diagrams contributing to
          $\Pp\Pp\to\mu^+\nu_\mu\Pe^+\nu_{\Pe}\Pj\Pj$ at
          $\order{\gs e^6}$.}
\label{diag:real}
\end{figure}

In this work, we compute the full NLO predictions to the process of Eq.\
\eqref{eq:process}, \ie all four NLO contributions of orders $\order{\alpha^7}$, $\order{\alphas\alpha^6}$, $\order{\alphas^2\alpha^5}$, and $\order{\alphas^3\alpha^4}$ are fully taken into account.
Our fixed-order off-shell calculation follows closely the one presented in
\citere{Biedermann:2017bss} for VBS with the same
final state. As explained
there in detail,  all virtual and real contributions are included and,
in particular, all off-shell and non-resonant contributions are taken
into account.
Some sample diagrams are shown in \reffis{diag:loop} and \ref{diag:real}.
In addition, we also compute the photon-induced contributions for all relevant orders [$\order{\alpha^7}$, $\order{\alphas\alpha^6}$, and $\order{\alphas^2\alpha^5}$].

One of the calculations presented in this work thus provides the full NLO and
off-shell predictions for $\Pp\Pp\to\mu^+\nu_\mu\Pe^+\nu_{\Pe}\Pj\Pj$
at the LHC in a tri-boson phase space in \refse{se:NLO}.
To consistently treat off-shell contributions, the complex-mass scheme~\cite{Denner:1999gp,Denner:2005fg,Denner:2006ic} is used throughout.

\subsection{Analysing the composition of the off-shell calculation}
\label{sec:composition}

In order to investigate the complexity and composition of the
complete off-shell calculation, both at LO and NLO, we decompose
the EW production mode taking into account all contributing on-shell
channels, \ie $\PW\PW\PW$ production, $\PW\PH$ production, $\PW\PZ$
production, and $\PW^+\PW^+\Pj\Pj$ production in the VBS approximation.
Besides discarding non-resonant contributions, this approach
also neglects interferences between different resonant channels.
The details of this approximation and its quality to capture
the full off-shell results are discussed in \refse{sec:onshell}.

In these on-shell calculations, the \PW, \PZ, and/or Higgs bosons
are produced on their mass shell, and their decays are calculated
using LO $1\to 2$ and $1\to 4$ decay matrix elements, including
full LO spin correlations \cite{Hoche:2014kca}.
We adjust the branching ratios to account for the respective LO
decay width and the measured total width used in the LO off-shell
calculation (see \refse{se:setup} for details).

Furthermore, we also study a decomposition of the fully off-shell
calculation in terms of $s$- and $t/u$-like partonic channels
of the underlying di-jet production topologies. Thereby we still treat
all internal propagators as fully off~shell. While the $s$~channel
comprises off-shell $\PW\PW\PW$, $\PW\PH$, and $\PW\PZ$ topologies
as well as their interferences, the $t/u$~channel consists of the
off-shell $\PW^+\PW^+$ VBS process (for details see the discussion
in \refse{se:s-t-channels}). This decomposition forms the basis
for our parton-shower matched predictions of the EW production mode.

\subsection{NLO QCD matched to parton shower with virtual \EW\ approximation}
\label{sec:NLOPS}

Besides the aforementioned fixed-order calculations, we provide
parton-shower matched predictions for the $\mu^+\nu_\mu\Pe^+\nu_{\Pe}\Pj\Pj$
final state, allowing one to model fully exclusive hadron-level events.
For the NLO QCD accurate parton-shower simulations with \Sherpa\
we separate the full process into its QCD and EW production modes,
corresponding to the $\order{\alpha^6}$ and $\order{\alpha_s^2\alpha^4}$
terms at LO, respectively. The interference contribution of
$\order{\alpha_s\alpha^5}$ is not included. 

In addition to the NLO QCD corrections, which are matched to the \Sherpa
dipole shower~\cite{Schumann:2007mg}, using the methods detailed in
\citere{Hoeche:2011fd}, for the EW production mode we furthermore
incorporate NLO EW corrections through the \EWvirt\ approximation
\cite{Kallweit:2015dum,Gutschow:2018tuk}, which, in particular, captures
the dominant effects in the high-energy limit~\cite{Brauer:2020kfv,
  Bothmann:2021led}. 

While for the EW production mode we use an incoherent decomposition
into pure $s$- and $t/u$-channel-like contributions, the QCD production
process is treated in full generality, including NLO QCD corrections
and shower evolution, as well as resolved final-state photon emissions
off the charged leptons in the YFS \cite{Yennie:1961ad} formalism.
However, for this channel we do not account for EW corrections, as these
cannot unambiguously be assigned to the QCD production mode (\cf \refse{se:mcatnlo}).

The calculational setup used and described here, based on the \Sherpa event
generator, is fully realisable in the computing frameworks of the LHC experiments,
such that corresponding particle-level simulations can be obtained and utilised in
future data analyses.

\subsection{Technical aspects and tools}
\label{se:tools}

For the full NLO computation at fixed order, the combination of codes \MoCaNLORecola has been used.
\mocanlo is a general NLO Monte Carlo program which has been shown to be particularly adapted for high-multiplicity processes such as $\Pp\Pp\to VV'\Pj\Pj$ at NLO QCD and EW accuracy~\cite{Biedermann:2016yds,Biedermann:2017bss,Denner:2019tmn,Denner:2020zit,Denner:2021hsa,Denner:2022pwc}.
To render the numerical integration of such complex processes
efficient, it uses multi-channel phase-space mappings as 
introduced in \citeres{Berends:1994pv,Denner:1999gp,Dittmaier:2002ap}.
To ensure the  convergence over the full
phase space  all relevant integration channels have to be included,
usually by introducing one integration channel for each kinematically
different Feynman diagram. Furthermore, for 
contributions like $\Pp\Pp\to \PW^+\PH(\to\PW^+\PW^-)$, with different
mutually exclusive resonance structures, a permutation of the order of
the generated resonances has been implemented in \mocanlo following
the ideas of \citere{Knippen:2019ojd}. Moreover, resonant contributions in the
dipole-subtraction terms of the real NLO corrections are taken care of
by tailored integration channels \cite{Denner:2002cg}. Both types of extra channels turned out to be particularly relevant for contributions involving a potentially
resonant Higgs boson and two potentially resonant W~bosons.
The infrared (IR) singularities arising from the real QCD or QED radiations are treated by
the dipole-subtraction method \cite{Catani:1996vz,Dittmaier:1999mb,Catani:2002hc,Phaf:2001gc}.
Note that for the present computation, we did not make use of a recent FKS-scheme~\cite{Frixione:1995ms} implementation~\cite{Denner:2023grl} in MoCaNLO.
For all amplitudes (either tree or loop ones), the matrix-element generator \Recola \cite{Actis:2016mpe,Actis:2012qn} has been used.
It employs the \collier library \cite{Denner:2014gla,Denner:2016kdg} for the numerical evaluation of the one-loop scalar \cite{'tHooft:1978xw,Beenakker:1988jr,Dittmaier:2003bc,Denner:2010tr}
and tensor integrals \cite{Passarino:1978jh,Denner:2002ii,Denner:2005nn}.

The NLO QCD computations of the EW production process in the on-shell
approximation and $s$- and $t/u$-channel decomposition, as well as all
parton-shower-matched NLO QCD calculations, including approximate NLO
EW corrections, have been performed within the \Sherpa framework~\cite{Bothmann:2019yzt,Gleisberg:2008ta}.
Tree-level amplitudes and phase-space integration channels are provided by
\Amegic~\cite{Krauss:2001iv} and for the real-emission subtraction by \Comix~\cite{Gleisberg:2008fv}.
All one-loop contributions are obtained from \Recola~\cite{Actis:2016mpe,Actis:2012qn} using a
general interface to \Sherpa~\cite{Biedermann:2017yoi}. Infrared QCD and QED singularities are
treated according to the dipole-subtraction formalism~\cite{Catani:1996vz,Dittmaier:1999mb,Catani:2002hc,Phaf:2001gc} with
the dedicated implementation in \Sherpa~\cite{Gleisberg:2007md,Schonherr:2017qcj}.
The full QCD NLO matrix elements get matched to the \Sherpa dipole shower~\cite{Schumann:2007mg}
based on the \MCatNLO formalism~\cite{Hoeche:2012yf}. NLO EW corrections are included in the virtual
approximation (see for example \citeres{Brauer:2020kfv,Bothmann:2021led} for a detailed
description). QED corrections to final-state charged leptons are included via the YFS
formalism \cite{Schonherr:2008av,Krauss:2018djz},
including the photon-splitting corrections of \citere{Flower:2022iew}. When considering the
on-shell approximation, the decays of massive particles produced on~shell in the
matrix-element calculation are treated by \Sherpa's internal decay-handler module, thereby
accounting for spin correlations and invoking a Breit--Wigner smearing for the
intermediate resonances~\cite{Hoche:2014kca}. To analyse events we make use of the \Rivet
package~\cite{Bierlich:2019rhm}.

\subsection{Setup}
\label{se:setup}

\subsubsection*{Numerical inputs}

The results obtained in the present work are for the LHC running at a
centre-of-mass energy of $\sqrt{s} = 13.6 \TeV$.  We use the
\texttt{NNPDF31\_nnlo\_as\_0118\_luxqed} parton distribution function
(PDF) set \cite{Bertone:2017bme}
via the \textsc{Lhapdf} interface \cite{Buckley:2014ana}. This PDF set
employs  $\alphas(M^2_\PZ)= 0.118$ for the strong coupling constant
and the method of 
\citere{Manohar:2016nzj} for the extraction of the photon distribution.
The renormalisation and factorisation scales have been set to
\begin{equation}
\label{eq:scale}
 \muR = \muF = m_{\rT, \Pj\Pj} + m_{\rT, \nu_\Pe \Pe^+} + m_{\rT, \nu_\mu \mu^+}\,,
\end{equation}
where $m_{\rT, ij} = \sqrt{m_{ij}^2 + p^2_{\rT, ij}}$ is the
transverse mass for the particle pair $i$ and $j$ with invariant mass $m_{ij}$. This scale is similar to the one 
used in \citere{Dittmaier:2019twg} for the calculation of NLO QCD and
EW corrections to triple-W-boson production with leptonic decays at the LHC.
In order to study the validity of the on-shell approximation we perform calculations
using an alternative fixed-scale definition that is in particular independent of the decay-product
kinematics, namely
\begin{equation}
\label{eq:fixed_scale}
 \muR = \muF = 3 M_{\PW}\,.
\end{equation}
In the off-shell computation, the following masses and widths are used:
\begin{alignat}{2}
                  \Mt   &=  173.0\GeV,       & \quad \quad \quad \Mb &= 0 \GeV,  \nonumber \\
                \MZOS &=  91.1876\GeV,      & \quad \quad \quad \GZOS &= 2.4952\GeV,  \nonumber \\
                \MWOS &=  80.379\GeV,       & \GWOS &= 2.085\GeV,  \nonumber \\
                M_{\rm H} &=  125.0\GeV,       &  \GH   &=  4.07 \times 10^{-3}\GeV.
\end{alignat}
Note that in the process \refeq{eq:process} no bottom or top quarks enter in tree-level amplitudes.
Since they only appear within loops, their respective widths are set to zero in the calculation.
The values of the Higgs-boson mass and width are taken from \citere{Heinemeyer:2013tqa}.
The pole masses and widths of the W and Z~bosons that are utilised in the numerical calculations are obtained from the measured on-shell (OS) values via~\cite{Bardin:1988xt}
\begin{equation}\label{eq:pole-mass-width}
        M_{\text{V}} = \frac{\MVOS}{\sqrt{1+(\GVOS/\MVOS)^2}}\;,\qquad
\Gamma_{\text{V}} = \frac{\GVOS}{\sqrt{1+(\GVOS/\MVOS)^2}}\;,
\end{equation}
with ${\text{V}}=\PW, \PZ$.
The \EW\ coupling is fixed through the $G_\mu$ scheme \cite{Denner:2000bj,Dittmaier:2001ay} upon
\begin{equation}
  \alpha = \frac{\sqrt{2}}{\pi} G_\mu \MW^2 \left( 1 - \frac{\MW^2}{\MZ^2} \right)  \qquad \text{and}  \qquad   \GF    = 1.16638\times 10^{-5}\GeV^{-2}\;.
\end{equation}

For the on-shell calculations in the narrow-width approximation,
the widths of the $\PW$, $\PZ$, and Higgs bosons are set to zero
in the matrix-element computations.
The produced bosons are subsequently decayed using the algorithm
of \citere{Hoche:2014kca}, preserving the LO spin correlations.
The branching ratios for each decay are determined by taking the ratio
of the LO decay width of the chosen decay channel over the total width
in the pole scheme of the decaying boson. The LO decay width of the
Higgs boson into 4 fermions is calculated with {\sc Prophecy4f}
\cite{Denner:2019fcr}, resulting in 
\begin{equation}
\Gamma(\PH\to\Pe^+\nu_\Pe\bar\Pu\Pd) + \Gamma(\PH\to\Pe^+\nu_\Pe\bar\Pc\Ps) = 0.06125\MeV.
\end{equation}
In addition, we adjust the kinematics of the intermediate resonance
according to a Breit--Wigner distribution using its pole mass and width
as input to mimic kinematic off-shell effects. This LO treatment of
the boson decays is also applied for on-shell NLO calculations.

\subsubsection*{Event selection}
\label{se:selection}

The event selection for the present calculation is a simplified version of the kinematic cuts used in the ATLAS measurement of \citere{ATLAS:2022xnu}.
The recombination of the QCD partons and photons (with $|y|<5$) is performed in two steps:
\begin{enumerate}
 \item First, photons and jets are recombined with the anti-$k_{\rm T}$ algorithm~\cite{Cacciari:2008gp} and a radius parameter $R=0.4$.
 \item Then, the non-clustered photons are recombined with the charged leptons with radius parameter $R=0.1$ using the Cambridge--Aachen algorithm~\cite{Dokshitzer:1997in}.
\end{enumerate}

In general, the experimental signature consists of two identified jets, two same-sign charged leptons (positron and anti-muon in the present case), and missing transverse energy (which is not explicitly required in the present selection).
In detail, the two dressed leptons must fulfil the conditions
\begin{align}
\label{eq:lepton}
\ptsub{\ell^+} >  20\GeV \qquad \textrm{and} \qquad |y_{\ell^+}| < 2.5,
\end{align}
where $y$ is the rapidity and $p_{\rT}$ the transverse momentum.
In addition, their invariant mass is constrained to be in the window
\begin{align}
\label{eq:lepton2}
40\GeV < m_{\ell^+\ell^+} < 400\GeV .
\end{align}
Finally, the identified anti-$k_{\rm T}$ QCD jets are defined through the two conditions
\begin{align}
\label{eq:jet}
 \ptsub{\Pj} >  20\GeV \qquad \textrm{and} \qquad |y_{\Pj}| < 4.5 .
\end{align}
From the list of all identified jets in an event [fulfilling Eq.~\eqref{eq:jet}], the two hardest jets (in transverse momentum) are further required to respect the conditions
\begin{align}
\label{eq:jet2}
 m_{\Pj\Pj} <  160\GeV \qquad \textrm{and} \qquad |\Delta y_{\Pj\Pj}| < 1.5 .
\end{align}

\section{Results}
\label{sec:results}

\subsection{LO contributions}

In this section, we first discuss results of the off-shell description of $\process$ at LO accuracy  at the LHC.
As explained in \refse{sec:features}, at LO the partonic process contains three different contributions of
order $\order{\alpha^6}$, $\order{\alphas\alpha^5}$, and $\order{\alphas^2\alpha^4}$, respectively.
The corresponding fiducial cross section, as well as those of the separate parts, are provided in \refta{tab:LO}.

\begin{table}
  \begin{center}
    \begin{tabular}{c||c|c|c||c}
     order\hl & $\order{\alpha^6}$ & $\order{\alphas\alpha^5}$ & $\order{\alphas^2\alpha^4}$ & sum \\
     \hline
     $\sigma_\text{LO} [\text{fb}]$\hl & $0.78549(9)$ & $0.00732(1)$ & $0.25925(3)$ & $1.05206(9)$ \\
     \hline
     $\sigma / \sigma_\text{LO}^\text{sum} [\%]$\hl & $74.7$ & $0.7$ & $24.6$ & $100$ \\
    \end{tabular}
  \end{center}
  \caption{\label{tab:LO}
  Cross sections at LO accuracy for $\process$ at the LHC for the
  three contributing orders and their sum using the dynamical scale
  defined by Eq.~\refeq{eq:scale}. The second line contains the absolute predictions,
  while the third one provides the relative numbers normalised to the
  sum of all contributions. 
    }
\end{table}

With the event selection defined in \refse{se:selection}, the EW contribution is about $75\%$ of the fiducial cross section.
As a comparison, in \citere{Biedermann:2017bss} where the same final state is computed in a VBS phase space, the EW part amounts to a bit more than $85\%$.
Typical VBS event selections require large invariant masses and large rapidity differences of the two jets to single out the EW component.
Taking advantage of the high-energy behaviour of VBS amplitudes over QCD-mediated ones allows one to obtain samples highly enriched in EW contributions.
On the other hand, the tri-boson phase space necessarily requires a low invariant mass for the two jets (of the order of the W-boson mass) where the QCD component is not suppressed~\cite{Ballestrero:2018anz} resulting in lower purity.
The interference contribution is small, below $1\%$.
Note that in the present case, in addition to being simply colour
suppressed, there is a further cancellation between partonic
channels. While the $t$--$u$ interference channels come with a
positive contribution, those involving $s$~channels are negative.
Combining all contributions, the full LO prediction reads
\begin{align}
 \sigma_{\rm LO}^{\rm sum} = 1.05206(9)^{+4.9\%}_{-3.8\%} \; \fb.
\end{align}
The subscript and superscript indicate the 7-point scale variation,
which amounts to taking the envelope of the predictions obtained by
scaling the renormalisation and factorisation scales defined in
Eq.~\eqref{eq:scale} by the factors $\left(\xi_{\rm F}, \xi_{\rm R}\right) \in \{ (1/2, 1/2) , (1/2, 1) , (1, 1/2) , (1, 1) , (1, 2) , (2, 1) , (2, 2) \}$.
This rather small scale dependence at LO is driven by the fact that the dominating EW contribution does not carry a renormalisation-scale dependence.

\subsection{On-shell approximations}
\label{sec:onshell}

In this section, we discuss the quality and implications of calculating the
production of $\mu^+\nu_\mu\Pe^+\nu_{\Pe}\Pj\Pj$ in the on-shell
approximation at LO for the EW production mode.
Because of the presence of multiple sequential and competing resonances
in the full process (see \reffi{diag:LO}),
we include four separate on-shell production channels at $\order{\alpha^6}$:
\begin{itemize}
\item[a)] $\PW^+\PW^+\PW^-$, where the $\PW^+$ bosons decay into
  electron/electron neutrino and muon/muon neutrino, while the
  $\PW^-$ boson decays hadronically.

\item[b)] $\PW^+\PH$, where the $\PW^+$ boson decays into
  electron/electron neutrino or muon/muon neutrino, while the
  Higgs boson decays into four fermions semi-leptonically, containing
  a muon/muon neutrino or electron/electron neutrino pair, respectively.
  The Higgs decay width into 4 fermions is calculated with {\sc Prophecy4f}
  \cite{Denner:2019fcr}.
  
\item[c)] $\PW^+\PZ$, where the $\PW^+$ decays into electron/electron
  neutrino or muon/muon neutrino, while the $\PZ$~boson decays into four
  fermions semi-leptonically, containing a muon/muon neutrino or 
  electron/electron neutrino pair, respectively. 
  The \PZ-boson decay is thereby approximated by the production of an on-shell
  W~boson (with subsequent decay) and two fermions.  Both cases of an
  on-shell W$^-$ and on-shell W$^+$ boson are added.

  \item[d)]  $\PW^+\PW^+$ production in the VBS topology. Here all
    $s$-channel Feynman diagrams, \ie diagrams where the incoming
    quarks are detached from the outgoing quarks, are discarded, while
    all $t$- and $u$-channel diagrams and interferences between them
    are retained.%
    \punctfootnote{
      This is at variance to the traditional
      VBS approximation, where also interferences between $t$-
      and $u$-channel diagrams are neglected.}
    This ensures that this category has no overlap with
    categories a--c.  The $\PW^+$ bosons decay into
    electron/electron neutrino and muon/muon neutrino.
\end{itemize}
At $\order{\alphas^2\alpha^4}$ and $\order{\alphas\alpha^5}$,
in the absence of other resonant channels, we include all contributions to
the complete $\PW^+\PW^+\Pj\Pj$ final state. As mentioned before, when 
investigating the quality of the on-shell approximation, a fixed
renormalisation and factorisation scale of $\muR=\muF=3M_\PW$ is used.

\subsubsection{On-shell approximations at LO}
\label{sec:onshell:LO}

\begin{table}
  \begin{center}
    \begin{tabular}{c||C||C||C|C|C|C}
      $\order{\alpha^6}$
      & off~shell\hl & on~shell & \multicolumn{4}{c}{on-shell subprocess} \\\cline{1-7}
      \multirow{2}{*}{Process} & \multirow{2}{*}{$\!\!\processfs$} &
      \multirow{2}{*}{sum} & \multirow{2}{*}{$\!\PW^+\PW^+\PW^-$} &
      \multirow{2}{*}{$\PW^+\PH$} & \multirow{2}{*}{$\PW^+\PZ$} &
      $\PW^+\PW^+$\hl \\
      & & & & & & VBS\\
      \hline
      $\sigma_\text{LO} [\text{fb}]$\hl &
      $0.7917$ & $0.7738$ &
      $0.4207$ & $0.3265$ &
      $5\cdot 10^{-7}$ & $0.0266$ \\\hline
      $\sigma / \sigma_\text{LO}^\text{off~shell} [\%]$\hl &
      $100$ & $97.7$ &
      $53.1$ & $41.2$ &
      $7\cdot 10^{-5}$ & $3.3$
    \end{tabular}\\[5mm]
    \begin{tabular}{c||C||Ccc||C||C}
      $\order{\alphas^2\alpha^4}$ & off~shell & on~shell &
      \hspace*{30pt} &
      $\order{\alphas\alpha^5}$ & off~shell & on~shell \\\cline{1-3}\cline{5-7}
      \multirow{2}{*}{Process} & \multirow{2}{*}{$\!\!\processfs$} &
      \multirow{2}{*}{$\PW^+\PW^+\Pj\Pj$} &&
      \multirow{2}{*}{Process} & \multirow{2}{*}{$\!\!\processfs$} &
      \multirow{2}{*}{$\PW^+\PW^+\Pj\Pj$} \\&&&&&&\\\cline{1-3}\cline{5-7}
      $\sigma_\text{LO} [\text{fb}]$\hl &
      $0.2912$ & $0.2938$ &&
      $\sigma_\text{LO} [\text{fb}]$\hl &
      $0.0071$ & $0.0074$ \\\cline{1-3}\cline{5-7}
      $\sigma / \sigma_\text{LO}^\text{off~shell} [\%]$\hl &
      $100$ & $100.9$ &&
      $\sigma / \sigma_\text{LO}^\text{off~shell} [\%]$\hl &
      $100$ & $104.2$
    \end{tabular}
  \end{center}
  \caption{\label{tab:LO-onshell-offshell}
    Cross sections for off-shell production and on-shell
    approximations for $\process$ at LO using $\mu_{\rm R}=\mu_{\rm
      F}=3\MW$ throughout. The statistical integration errors are at
    most one in the last digits shown.}
\end{table}
In \refta{tab:LO-onshell-offshell}, the cross sections for all
contributing on-shell processes are compared to the off-shell one at
LO accuracy.  We notice that the on-shell approximation reproduces the full
off-shell results for the contributions of order $\order{\alpha^6}$
within less than $3\%$, which is in agreement with the expected precision of the
on-shell approximation.
Most strikingly, tri-boson production makes up only $53\%$ of the full
$\order{\alpha^6}$ off-shell cross section.
On the other hand, $\PW\PH$ production amounts to $41\%$.
Thus, for this phase space, which is supposedly a tri-boson one, almost half of the cross section actually results from the Higgs-strahlung process.
We note that the contribution of VBS production is at a level of few
per cent only, and the $\PW\PZ$ contribution is negligible.

For the second largest contribution to $\process$, of order $\order{\alphas^2\alpha^4}$,
the on-shell approximation reproduces the full off-shell result very well, yielding
a cross section that is larger by only $0.9\%$. For 
the interference contributions of order $\order{\alphas\alpha^5}$,
which amount to less than one per cent of the cross section, the
difference between on-shell and off-shell calculation is $4\%$.

\begin{figure}
  \centering
  \begin{subfigure}{0.47\textwidth}
    \includegraphics[width=\textwidth]{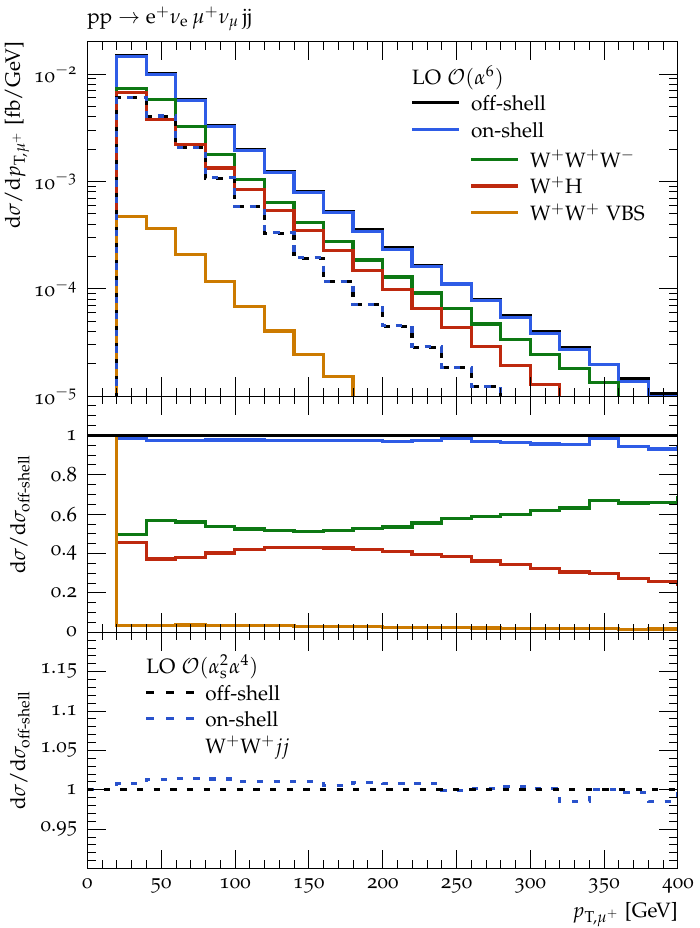}
  \end{subfigure}
  \begin{subfigure}{0.47\textwidth}
    \includegraphics[width=\textwidth]{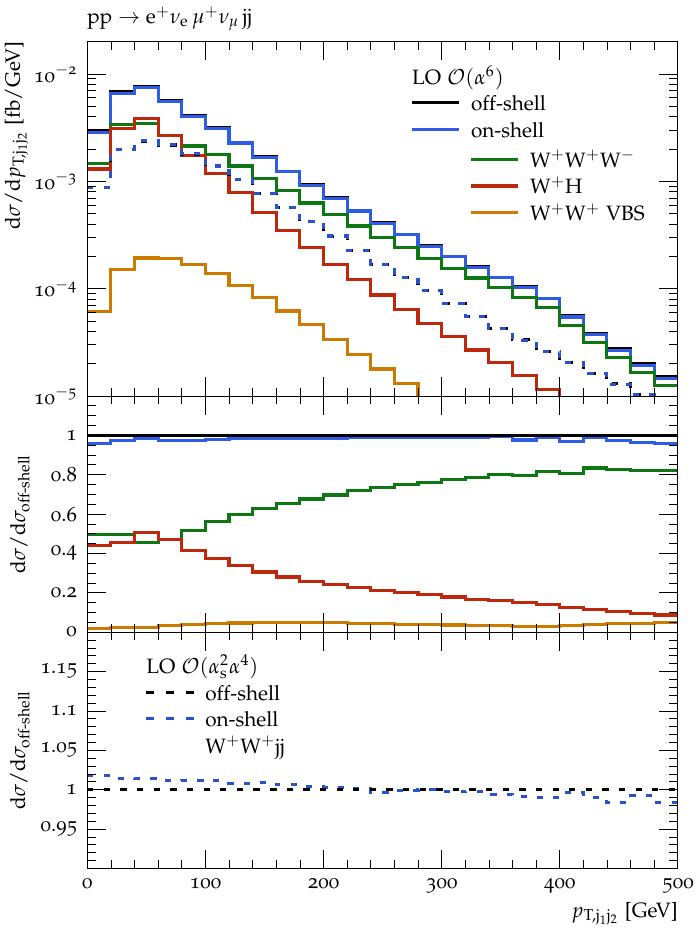}
  \end{subfigure}
  \\
  \begin{subfigure}{0.47\textwidth}
    \includegraphics[width=\textwidth]{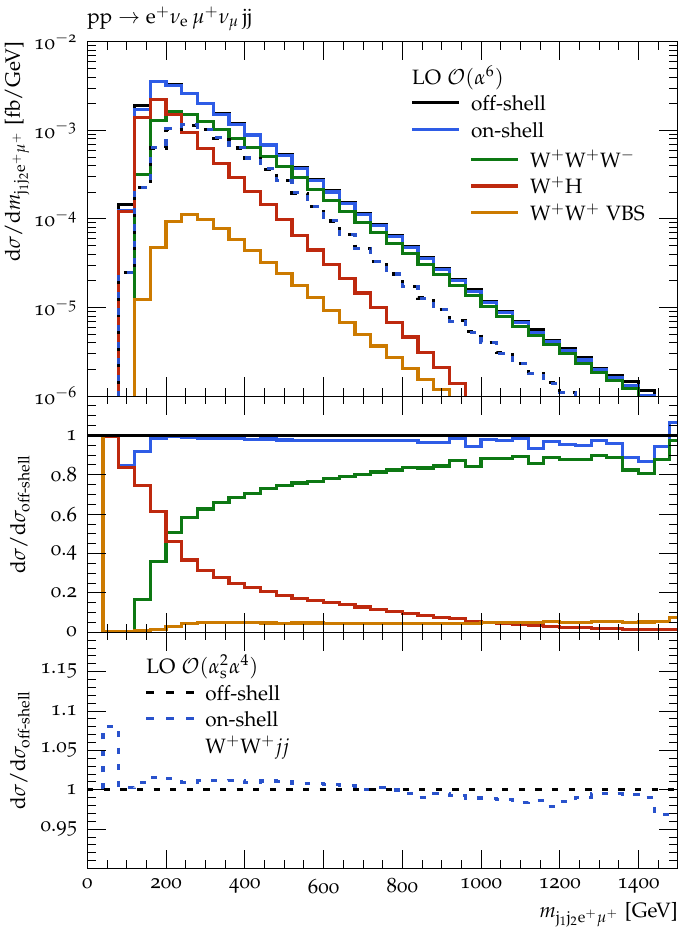}
  \end{subfigure}
  \begin{subfigure}{0.47\textwidth}
    \includegraphics[width=\textwidth]{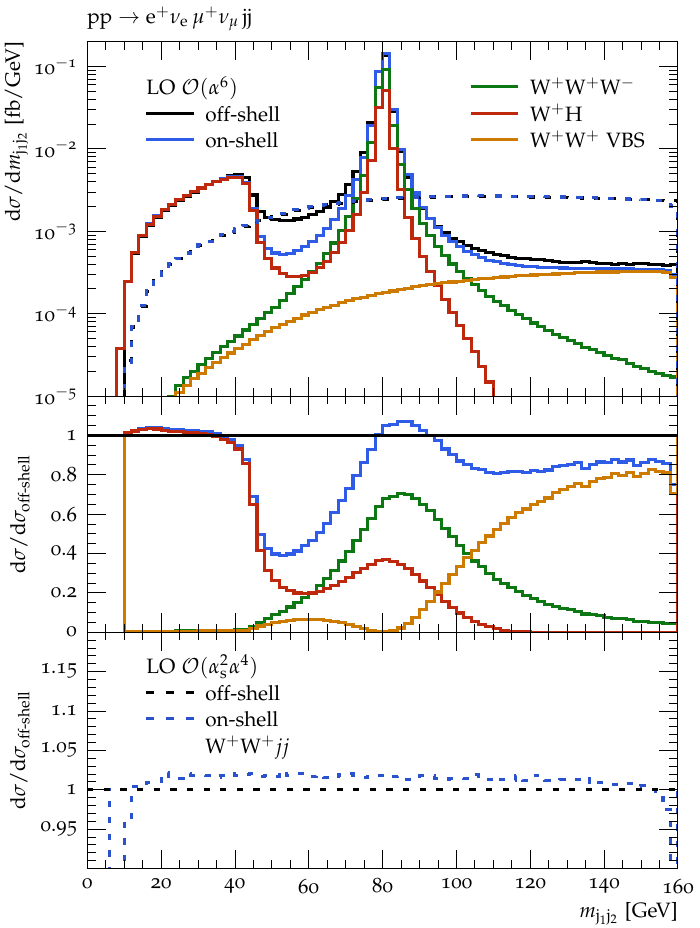}
  \end{subfigure}
  \caption{Differential distributions for the full off-shell process
  $\process$ and its relevant on-shell sub-contributions at LO,
    using $\mu_{\rm R}=\mu_{\rm F}=3\MW$. 
  The observables are: the transverse momentum of the anti-muon (top
  left), the transverse momentum of the two jets (top right), the
  invariant mass of the two jets and two charged leptons (bottom
  left), and the invariant mass of the two jets (bottom right).
    \label{fig:onshell-off-shell-LO-pT-m}
  }
\end{figure}
\begin{figure}
  \centering
  \begin{subfigure}{0.47\textwidth}
    \includegraphics[width=\textwidth]{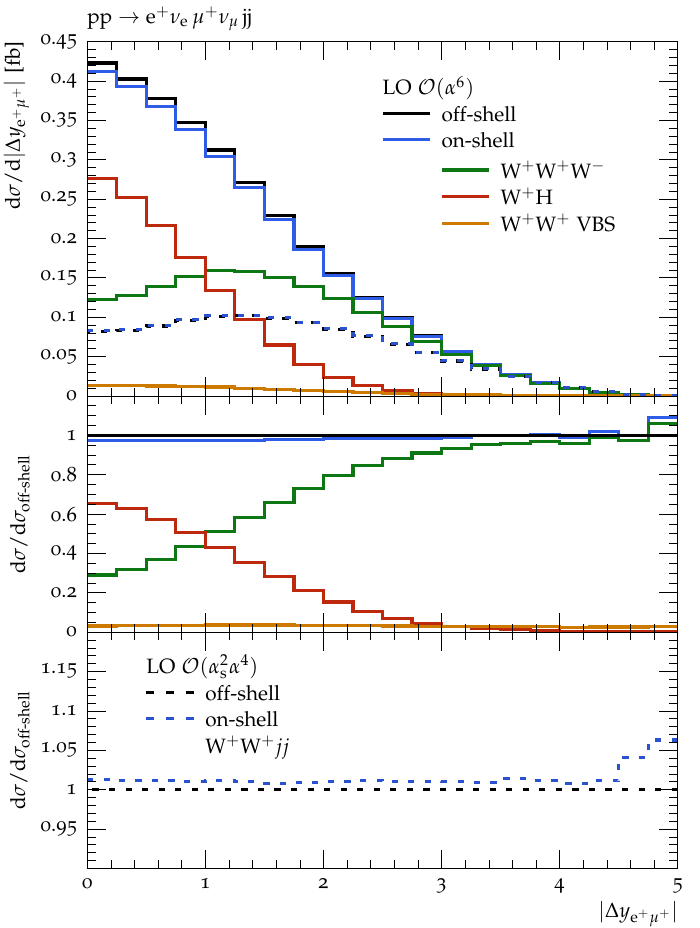}
  \end{subfigure}
  \begin{subfigure}{0.47\textwidth}
    \includegraphics[width=\textwidth]{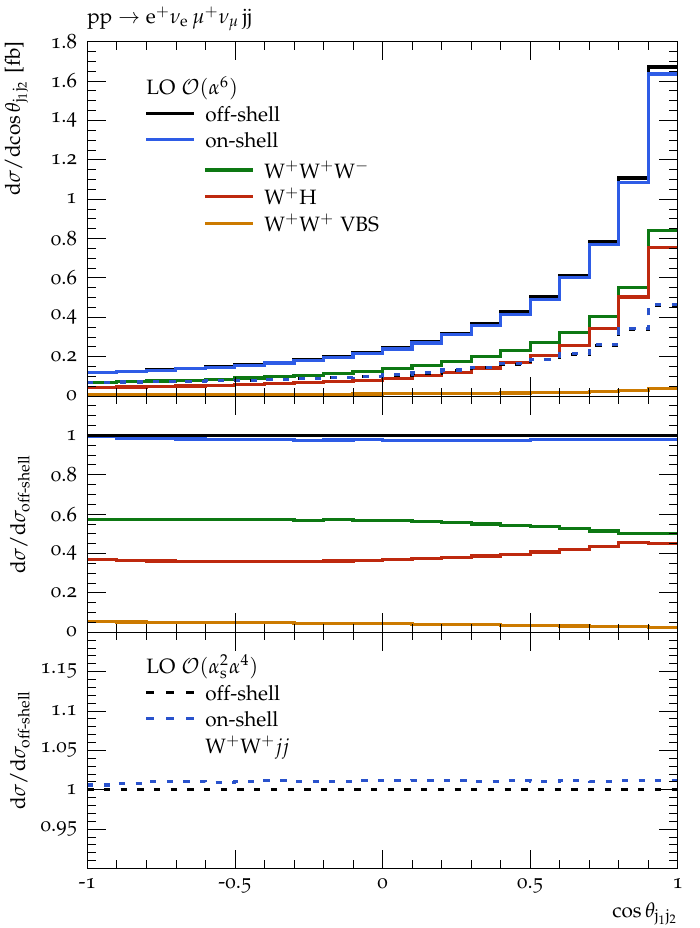}
  \end{subfigure}
  \caption{
  Differential distributions for the full off-shell  process
  $\process$ and its relevant on-shell sub-contributions at LO, using
  $\mu_{\rm R}=\mu_{\rm F}=3\MW$.
  The observables are: the modulus of the rapidity difference between the two charged leptons (left) and the cosine of the angle between the two jets (right).
    \label{fig:onshell-off-shell-LO-dy-ct}
  }
\end{figure}
Figures \ref{fig:onshell-off-shell-LO-pT-m} and \ref{fig:onshell-off-shell-LO-dy-ct} inspect the same aspect at the differential level.
The upper panels contain absolute predictions at order $\order{\alpha^6}$ for the off-shell process,
for the contributing on-shell processes (apart from the completely negligible
$\PW^+\PZ$ contribution) and their sum. The middle
panels show the ratio of the on-shell contributions to the off-shell
cross section at order $\order{\alpha^6}$ and the lower panels the
ratio at order $\order{\alphas^2\alpha^4}$. We do not investigate the
small interference contributions at the differential level.

While the inclusive cross section of the off-shell calculation is
reproduced on the level of $2\%$, differential distributions show deviations
of up to $20\%$ in various phase-space regions.
The differences are largest in the invariant-mass distribution of the
two jets where they reach $60\%$ for $m_{\Pj_1\Pj_2}\approx50\GeV$.
They are driven by  missing off-shell
contributions  as well as missing
interference effects between the various on-shell channels which have
different resonance structures. 
Note also that the on-shell
approximation overestimates the cross section for $m_{\Pj_1\Pj_2}$ below
$40\GeV$ and close to the \PW-boson mass.
In the distributions in $p_{\rT,\mu^+}$, $p_{\rT,\Pj_1\Pj_2}$, and
$m_{\Pj_1\Pj_2\Pe^+\mu^+}$, the differences between on-shell
approximation and off-shell calculation somewhat grow with the variables
values.

The two transverse-momentum distributions
(top row in \reffi{fig:onshell-off-shell-LO-pT-m}) display a similar
qualitative behaviour with $\PW\PH$ production contributing up to
$50\%$ of the cross section at low transverse momenta while becoming
small towards high $p_{\rT}$ where tri-boson production is dominant.
For the distribution in the invariant mass of the two jets and two
charged leptons (bottom left in \reffi{fig:onshell-off-shell-LO-pT-m}) $\PW\PH$ production dominates for low invariant masses
(below the triple-\PW threshold) but becomes negligible for large
invariant masses. The fraction of $\PW^+\PW^+$ VBS is roughly constant
except for small invariant masses.
As mentioned previously, the distribution in the invariant
mass of the jet pair (bottom right in
\reffi{fig:onshell-off-shell-LO-pT-m}) is characterised by several
resonance structures that interfere strongly. 
For $\MH-\MW \lsim  m_{\Pj_1\Pj_2} \lsim \MW$, all on-shell production
modes are suppressed (with at least one off-shell \PW or Higgs boson).
On the other hand, for $m_{\Pj_1\Pj_2}\gsim120\GeV$, VBS production is
becoming dominant as expected for high di-jet masses.
We stress that this picture holds only true at LO while higher-order QCD corrections significantly modify it.
In particular, tri-boson contributions are still very large
for $m_{\Pj_1\Pj_2}>100\GeV$ at NLO QCD as shown later (or as observed in \citeres{Ballestrero:2018anz,Denner:2020zit}).

Turning to angular distributions in
\reffi{fig:onshell-off-shell-LO-dy-ct}, $\PW\PH$ production is
dominant (at the level of $60\%$) at low rapidity difference between
the charged leptons.  At large rapidity difference tri-boson
production is overwhelming, reaching more than $90\%$ at $|\Delta
y_{\Pe^+\mu^+}|\gtrsim 3$.  On the other hand, the distribution in the cosine
of the angle between the two jets does not show strong variations in
the composition.

The difference between off-shell and on-shell calculations at
$\order{\alphas^2\alpha^4}$ remains below $2\%$ except in phase-space
regions with very small cross sections.

\subsubsection{On-shell approximations at NLO QCD}
\label{sec:onshell:NLOQCD}

\begin{table}
  \begin{center}
    \begin{tabular}{c||C||C||C|C|C|C}
      $\order{\alpha^6+\alphas\alpha^6}$
      & off~shell\hl & on~shell & \multicolumn{4}{c}{on-shell subprocess} \\\cline{1-7}
      \multirow{2}{*}{Process} & \multirow{2}{*}{$\!\!\processfs$} &
      \multirow{2}{*}{sum} & \multirow{2}{*}{$\!\PW^+\PW^+\PW^-$} &
      \multirow{2}{*}{$\PW^+\PH$} & \multirow{2}{*}{$\PW^+\PZ$} &
      $\PW^+\PW^+$\hl \\
      & & & & & & VBS\\
      \hline
      $\sigma_\text{NLO} [\text{fb}]$\hl &
      $1.123$ & $1.080$ &
      $0.542$ & $0.451$ &
      $1.8\cdot 10^{-6}$ & $0.086$ \\\hline
      $K_\text{NLO}$\hl &
      $1.42$ & $1.40$ &
      $1.29$ & $1.38$ &
       & $3.25$ \\\hline
      $\sigma / \sigma_\text{NLO}^\text{off~shell} [\%]$\hl &
      $100$ & $96.2$ &
      $48.3$ & $40.2$ &
      $1.6\cdot 10^{-6}$ & $7.7$
    \end{tabular}\\[5mm]
    \begin{tabular}{c||C||CccCC}
      $\order{\alphas^2\alpha^4+\alphas^3\alpha^4}$ & off~shell & on~shell &
      \hspace*{30pt} &
      & & \\\cline{1-3}
      \multirow{2}{*}{Process} & \multirow{2}{*}{$\!\!\processfs$} &
      \multirow{2}{*}{$\PW^+\PW^+\Pj\Pj$} &&
      & & \\&&&&&&\\\cline{1-3}
      $\sigma_\text{NLO} [\text{fb}]$\hl &
      $0.525$ & $0.520$ &&
       &
       &  \\\cline{1-3}
      $K_\text{NLO} [\text{fb}]$\hl &
      $1.80$ & $1.77$ &&
       &
       &  \\\cline{1-3}
      $\sigma / \sigma_\text{NLO}^\text{off~shell} [\%]$\hl &
      $100$ & $99.0$ && \phantom{$\sigma / \sigma_\text{NLO}^\text{off~shell} [\%]$}\hl
       &
       &
    \end{tabular}
    \hfill
  \end{center}
  \caption{\label{tab:NLO-onshell-offshell}
    Cross sections for off-shell production and on-shell
    approximations for $\process$ at NLO using $\mu_{\rm R}=\mu_{\rm
      F}=3\MW$ throughout. The statistical integration errors are at
    most one in the last digits shown.
    Please note that, contrary to all other NLO cross sections,
    the off-shell $\order{\alphas\alpha^6}$ process does not
    only comprise QCD-type corrections to its listed LO process but
    also EW-type corrections to a different Born process.
    The NLO $K$~factor is defined as the ratio of the quoted LO and
    NLO cross sections.
  }
\end{table}
We turn to the discussion of the quality of the on-shell
approximations, as defined in the beginning of this section, at NLO
QCD. We start by comparing fiducial off-shell and on-shell cross
sections in \refta{tab:NLO-onshell-offshell}. While the $K$~factors
are almost equal in both calculations, the difference between both
predictions rises to $4\%$ at NLO. This can be explained by the fact
that we do not include NLO QCD corrections to the decays of the
$\PW$~boson and the Higgs boson. For an inclusive $\PW$-boson decay these corrections
amount to $\alphas/\pi\approx 3\%$.  
The $K$~factors for the on-shell subprocesses
$\PW^+\PW^+\PW^-$ and $\PW^+\PH$ are in reasonable agreement with
literature results~\cite{Campanario:2008yg,Binoth:2008kt,Ferrera:2011bk}.
The contribution of $\PW^+\PW^+$ VBS is increased
by more than a factor of three. This can be explained as follows:
At LO the VBS process is suppressed by the cuts \refeq{eq:jet2}.
At NLO an additional radiated gluon can play the role of one leading
jet, allowing the two quark jets to have a large pair invariant mass
and rapidity separation and thus being in a phase-space region where
VBS is enhanced. At orders
$\order{\alphas^2\alpha^4+\alphas^3\alpha^4}$ the
difference between on-shell and off-shell cross sections is one per
cent, and the $K$~factors agree at this level.

\begin{figure}
  \centering
  \begin{subfigure}{0.47\textwidth}
    \includegraphics[width=\textwidth]{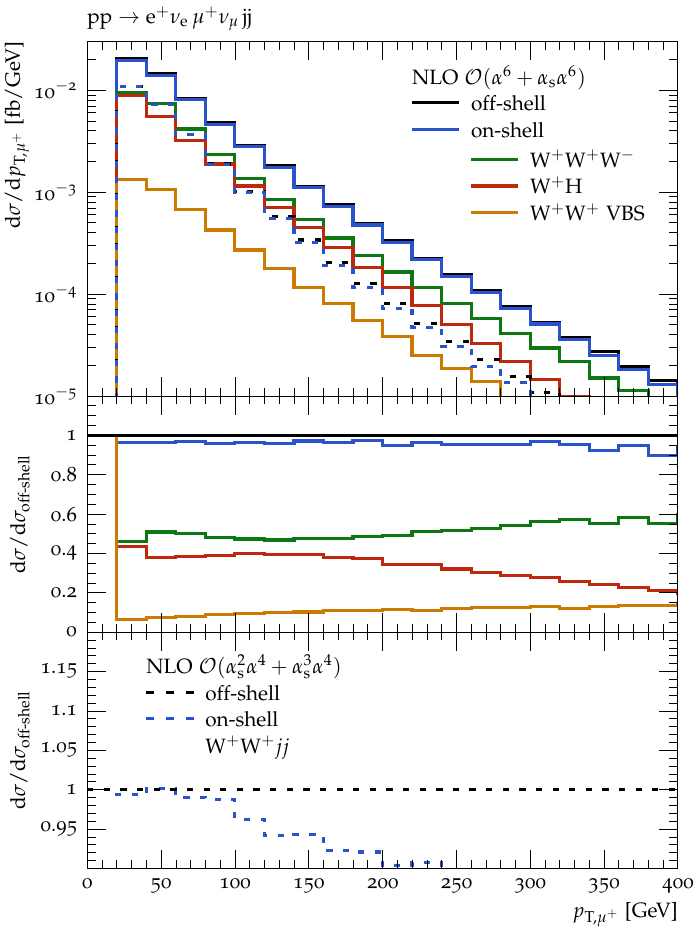}
  \end{subfigure}
  \begin{subfigure}{0.47\textwidth}
    \includegraphics[width=\textwidth]{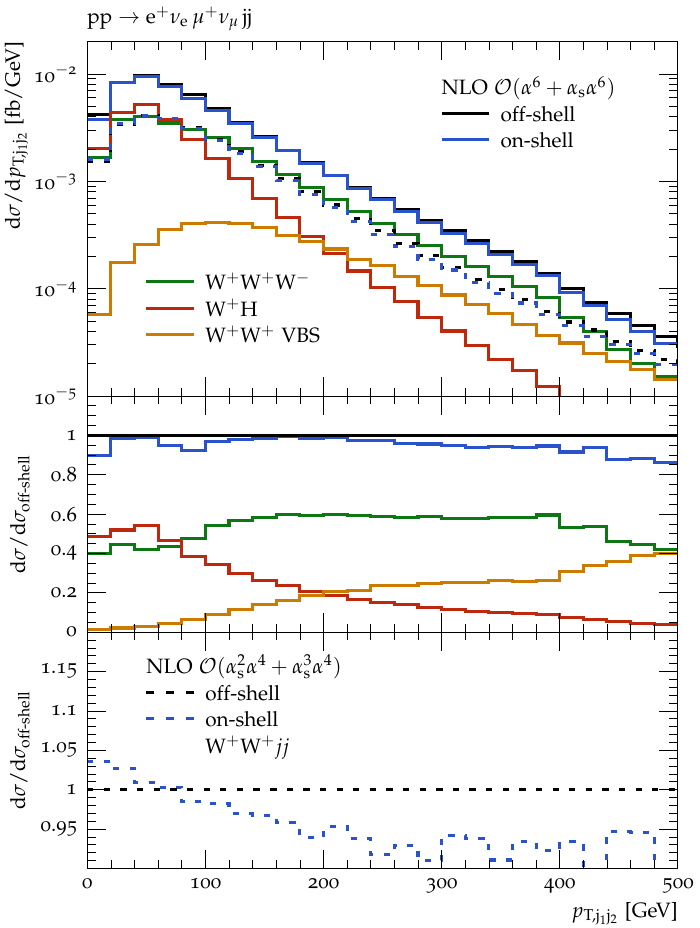}
  \end{subfigure}
  \\
  \begin{subfigure}{0.47\textwidth}
    \includegraphics[width=\textwidth]{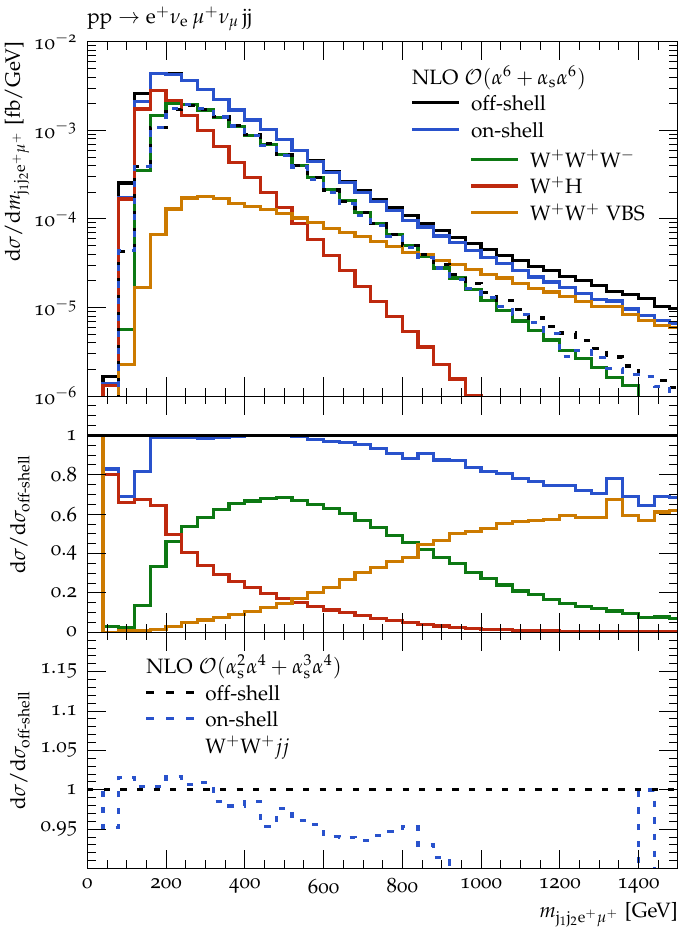}
  \end{subfigure}
  \begin{subfigure}{0.47\textwidth}
    \includegraphics[width=\textwidth]{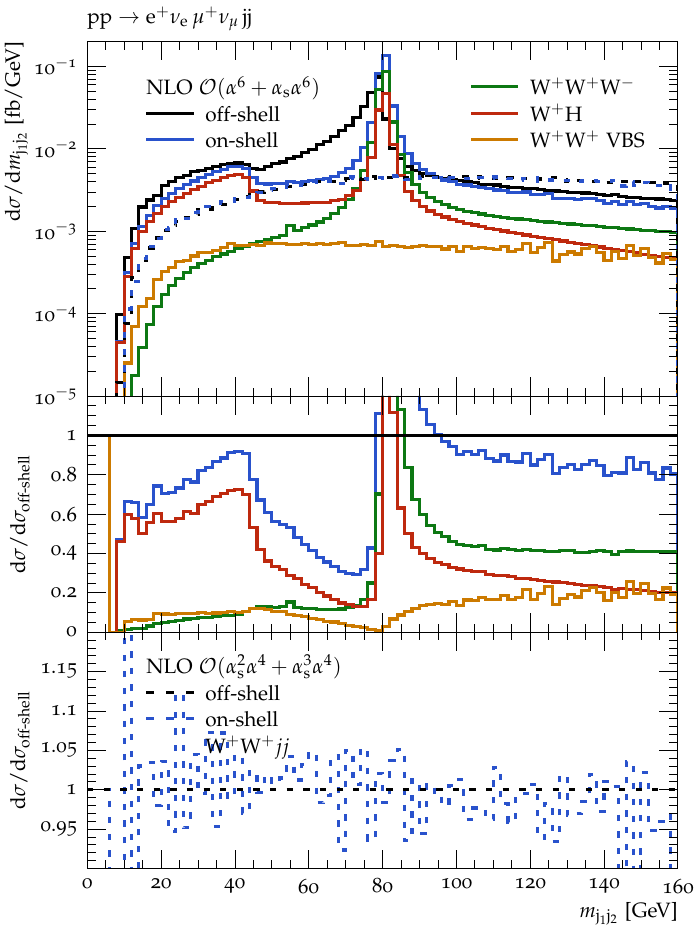}
  \end{subfigure}
  \caption{Differential distributions for the full off-shell process
  $\process(+\Pj)$ and its relevant on-shell sub-contributions at NLO QCD,
    using $\mu_{\rm R}=\mu_{\rm F}=3\MW$.
  The observables are: the transverse momentum of the anti-muon (top
  left), the transverse momentum of the two jets (top right), the
  invariant mass of the two jets and two charged leptons (bottom
  left), and the invariant mass of the two jets (bottom right).
    \label{fig:onshell-off-shell-NLO-pT-m}
  }
\end{figure}

\begin{figure}
  \centering
  \begin{subfigure}{0.47\textwidth}
    \includegraphics[width=\textwidth]{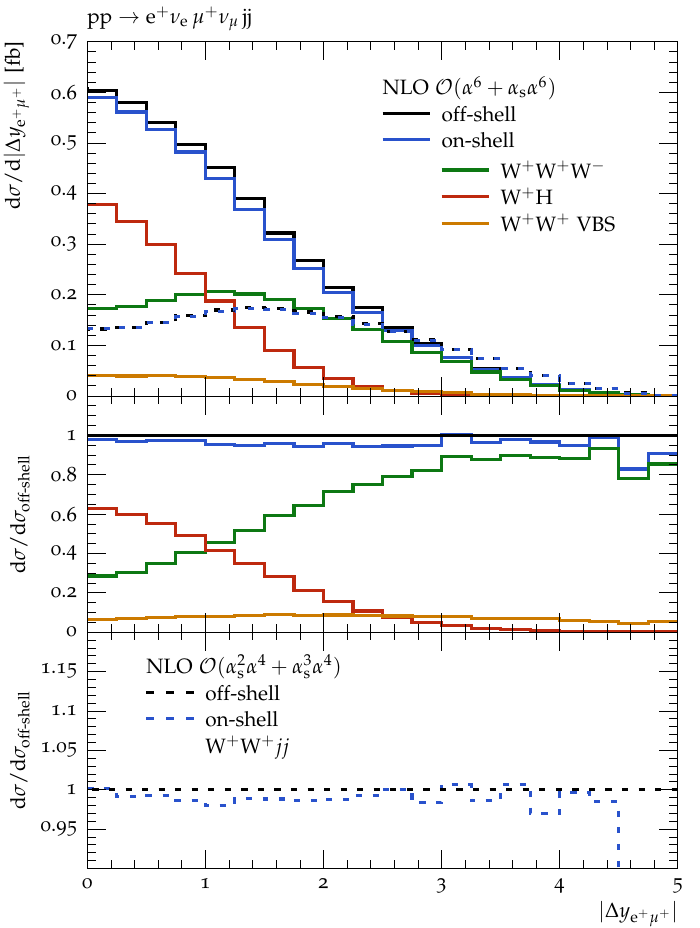}
  \end{subfigure}
  \begin{subfigure}{0.47\textwidth}
    \includegraphics[width=\textwidth]{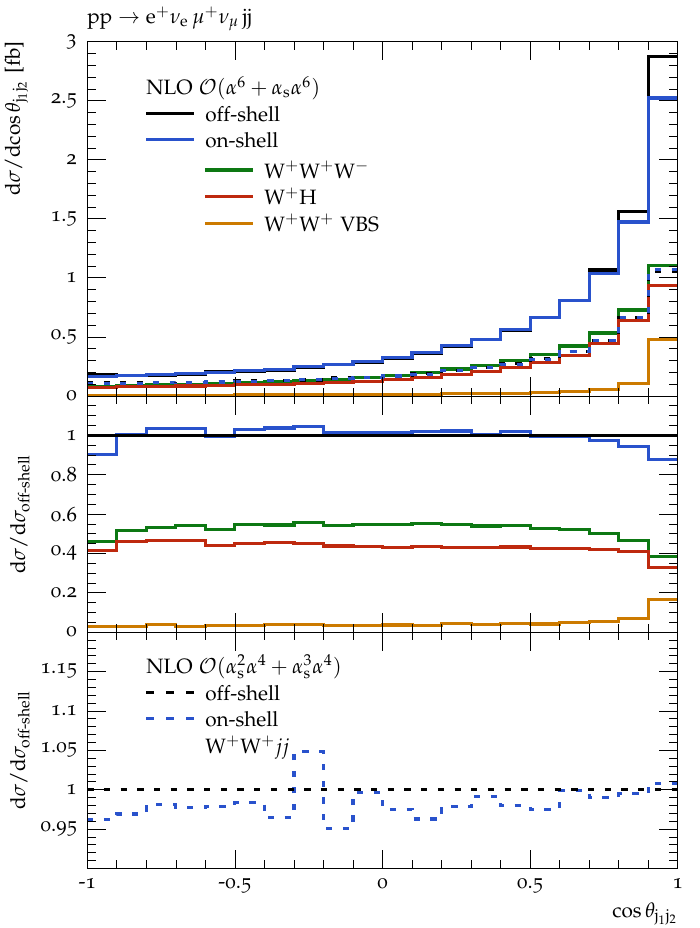}
  \end{subfigure}
  \caption{
Differential distributions for the full off-shell process
  $\process(+\Pj)$ and its relevant on-shell sub-contributions at NLO QCD,
    using $\mu_{\rm R}=\mu_{\rm F}=3\MW$.
  The observables are: the modulus of the rapidity difference between the two charged leptons (left) and the cosine of the angle between the two jets (right).
    \label{fig:onshell-off-shell-NLO-dy-ct}
  }
\end{figure}

In \reffis{fig:onshell-off-shell-NLO-pT-m} and
\ref{fig:onshell-off-shell-NLO-dy-ct} we present an analysis of the
on-shell approximation at $\order{\alpha^6+\alphas\alpha^6}$
and $\order{\alphas^2\alpha^4+\alphas^3\alpha^4}$ at the
differential level. We consider the same distributions as in
\refse{sec:onshell:LO} and use the same layout as in \reffis{fig:onshell-off-shell-LO-pT-m} and
\ref{fig:onshell-off-shell-LO-dy-ct}. 

At $\order{\alpha^6+\alphas\alpha^6}$ 
the on-shell approximation is considerably enhanced for values of
$m_{\Pj_1\Pj_2}$ close to and slightly above $\MW$. This results from
contributions where one of the hardest jets is a gluon allowing the
invariant mass of the $\PW^-$-boson decay jet pair to be close to the
resonance while $m_{\Pj_1\Pj_2}$ is larger. Obviously, this
enhancement mechanism does not apply to $\PW^+\PW^+$ VBS.
On the other hand, the on-shell approximation is strongly
suppressed in the range $\MH-\MW\lsim m_{\Pj_1\Pj_2}\lsim\MW$. While
this region is filled in the off-shell calculation by the usual
redistribution of events with real final-state 
radiation, this contribution is missing in our on-shell calculation
that does not include QCD corrections to the bosons' decays. 
A similar though smaller effect is seen near the peak at
$m_{\Pj_1\Pj_2}=\MH-\MW$. In the distributions in $p_{\rT,\mu^+}$, $p_{\rT,\Pj_1\Pj_2}$, and
$m_{\Pj_1\Pj_2\Pe^+\mu^+}$, the differences between on-shell
approximation and off-shell calculation grow stronger with the
variables than at LO. This behaviour is even more pronounced at
$\order{\alphas^2\alpha^4+\alphas^3\alpha^4}$, where the difference
exceeds $10\%$ in the tails. We attribute these differences to
contributions of off-shell diagrams that are not present in an
on-shell approximation. Similar effects have, for instance, been found in
high-energy tails in $\PW^+\PW^-$ di-boson production when comparing the
off-shell calculation with the double-pole approximation \cite{Biedermann:2016guo}.

Looking at individual on-shell contributions we observe the following:
The NLO QCD corrections raise the contribution of WH production with
respect to WWW production as compared to the LO ratio. The
contributions of $\PW^+\PW^+$ VBS are enhanced for large $p_{\rT\mu^+}$,  and
$m_{\Pj_1\Pj_2}$ close to and below $\MW$ and, in
particular for large $p_{\rT\Pj_1\Pj_2}$ and large
$m_{\Pj_1\Pj_2\Pe^+\mu^+}$. The mechanism of this enhancement is
the same as for the NLO corrections to the full process explained in
\refse{se:NLO}. Note that for large $m_{\Pj_1\Pj_2}$, the fraction
of $\PW^+\PW^+$ VBS is actually reduced.

\subsubsection[\texorpdfstring{$s$- and $t$-channel}{s- and t-channel} di-jet production modes]
{\texorpdfstring{$\boldsymbol{s}$- and $\boldsymbol{t}$-channel}{s- and t-channel} di-jet production modes}
\label{se:s-t-channels}

While the decomposition of $\mu^+\nu_\mu\Pe^+\nu_{\Pe}\Pj\Pj$
production into its on-shell channels is useful to understand
the contributions of the underlying topologies, it inherently
limits the accuracy of the theoretical predictions
in our fiducial region.
Hence, if the production of the $\mu^+\nu_\mu\Pe^+\nu_{\Pe}\Pj\Pj$
signature is to be understood on the per-cent level this
picture has to be abandoned.
It is still useful, however, to distinguish between $s$- and
$t$- (and/or $u$-) channel topologies regarding the production of
the di-jet system.
In particular, this distinction converts the NLO corrections of
$\order{\alphas\alpha^6}$ into pure QCD corrections to the
respective $\order{\alpha^6}$ Born process, which allows us to
employ the parton-shower matching procedure available in \Sherpa
without modification (see \refse{se:mcatnlo}).

While most quark-induced partonic processes comprise either $s$ or $t/u$~channels,
and therefore such a separation is straight-forward, the 
partonic processes  $\Pu\bar\Pd/\bar\Pd\Pu\to \mu^+ \nu_\mu\Pe^+
\nu_\Pe \Pd \bar\Pu$ 
contain both (see \refta{tab:EW-corr-by-channel}).
We incoherently decompose these into their $s$- and $t/u$-channel contributions,
thereby substituting the full squared matrix element by the sum of corresponding
pure $s$- and $t/u$-channel squared matrix elements.
For example, to reduce the matrix element of the partonic process
$\Pu\bar{\Pd}\to\mu^+\nu_\mu\Pe^+\nu_\Pe\Pd\bar{\Pu}$ to its
$s$-channel contribution it is calculated by using the
$\Pu\bar{\Pd}\to\mu^+\nu_\mu\Pe^+\nu_\Pe\Ps\bar{\Pc}$ amplitude
instead, while using the
$\Pu\bar{\Ps}\to\mu^+\nu_\mu\Pe^+\nu_\Pe\Pd\bar{\Pc}$ amplitude
allows to calculate its $t$-channel part.
Its original PDF is kept irrespective of a possible change of
parton flavour in the stand-in matrix element.
In this way, the separation is gauge invariant and all topologies
are accounted for on a diagrammatic level at LO.%
\punctfootnote{Note that such a splitting based on partonic processes is
  not possible at NLO owing to
  the appearance of quark--gluon channels that involve diagrams with
  both $s$-channel and $t$-channel vector-boson propagators.
  Hence, for these partonic processes, we rely on a diagram selection
  only. In the $s$-channel contribution we include diagrams
  featuring no EW $t/u$-channel propagator (setting the \Sherpa
  parameter {\tt{Max\_N\_TChannels=0}}), whereas for the $t$-channel
  mode we demand diagrams to have at least one such propagator,
  imposed through the setting {\tt{Min\_N\_TChannels=0}}.
  We employ this separation for calculating both the NLO QCD fixed-order cross sections of this
  subsection and the parton-shower-matched results of \refse{se:mcatnlo}.
}
Of course, interferences between $s$ and $t/u$~channels are
omitted in this approach.
They, however, turn out to be negligible in most of the considered phase space.
In terms of included on-shell equivalents, the $s$~channel
thus contains the $\PW\PW\PW$, $\PW\PH$, and $\PW\PZ$ processes
as well as all their respective interferences and off-shell
contributions. Conversely, the such defined $t/u$~channel corresponds
to the VBS process defined above.

\begin{table}
  \begin{center}
    \begin{tabular}{c||C||C||C|C}
      $\order{\alpha^6+\alphas\alpha^6}$\hl
      & coherent & incoherent & \multicolumn{2}{c}{subchannels} \\\cline{1-5}
      \multirow{2}{*}{Process} & \multirow{2}{*}{$|s+t/u|^2$} &
      \multirow{2}{*}{\!\!$|s|^2+|t/u|^2$} & \multirow{2}{*}{$|s|^2$} &
      \multirow{2}{*}{$|t/u|^2$} \\
      & & & & \\
      \hline
      $\sigma_\text{LO} [\text{fb}]$\hl &
      $0.7855$ & $0.7841$ &
      $0.7576$ & $0.0265$ \\\hline
      $\sigma_\text{NLO} [\text{fb}]$\hl &
      $1.091$ & $1.084$ &
      $0.998$ & $0.086$ \\\hline
      $K_\text{NLO}$\hl &
      $1.39$ & $1.38$ &
      $1.32$ & $3.3$ \\\hline
      $\sigma / \sigma_\text{LO}^\text{off~shell} [\%]$\hl &
      $100$ & $99.8$ &
      $96.5$ & $3.3$ \\\hline
      $\sigma / \sigma_\text{NLO}^\text{off~shell} [\%]$\hl &
      $100$ & $99.4$ &
      $91.5$ & $7.9$
    \end{tabular}
  \end{center}
  \caption{
    Cross sections for off-shell $s$- and $t/u$-channel
    approximations as well as their coherent and incoherent sum
    at LO and NLO QCD with $\muR=\muF=m_{\rT, \Pj\Pj} + m_{\rT, \nu_\Pe \Pe^+} + m_{\rT, \nu_\mu \mu^+}$.
    The statistical integration errors are at
    most one in the last digits shown.
    Please note that, contrary to all other NLO cross sections,
    the off-shell coherent sum of $s$ and $t/u$ channels at
    $\order{\alphas\alpha^6}$ does not
    only comprise QCD-type corrections to its listed LO process but
    also EW-type corrections to a different Born process.
    The NLO $K$~factor is defined as the ratio of the quoted LO and
    NLO cross sections.
    \label{tab:NLO-s-t-channel}
  }
\end{table}
Results for the fiducial cross sections at $\order{\alpha^6}$
and including NLO QCD corrections are shown in
\refta{tab:NLO-s-t-channel}. The difference between the off-shell
calculation (coherent) and the incoherent sum of $s$-channel and
$t$-channel contributions amounts to only $0.2\%$ at LO and $0.6\%$ at
NLO. The corresponding $K$ factors are practically identical.
Note also that the results for the off-shell $t/u$-channel
contributions in \refta{tab:NLO-s-t-channel} are almost identical to
those of the on-shell calculation in \reftas{tab:LO-onshell-offshell}
and \ref{tab:NLO-onshell-offshell}.
The differences between the off-shell $s$-channel contributions in
\refta{tab:NLO-s-t-channel} and the sum of the corresponding terms in 
\reftas{tab:LO-onshell-offshell}
and \ref{tab:NLO-onshell-offshell} are at the level of $1\%$.

\begin{figure}
  \centering
  \begin{subfigure}{0.47\textwidth}
    \includegraphics[width=\textwidth]{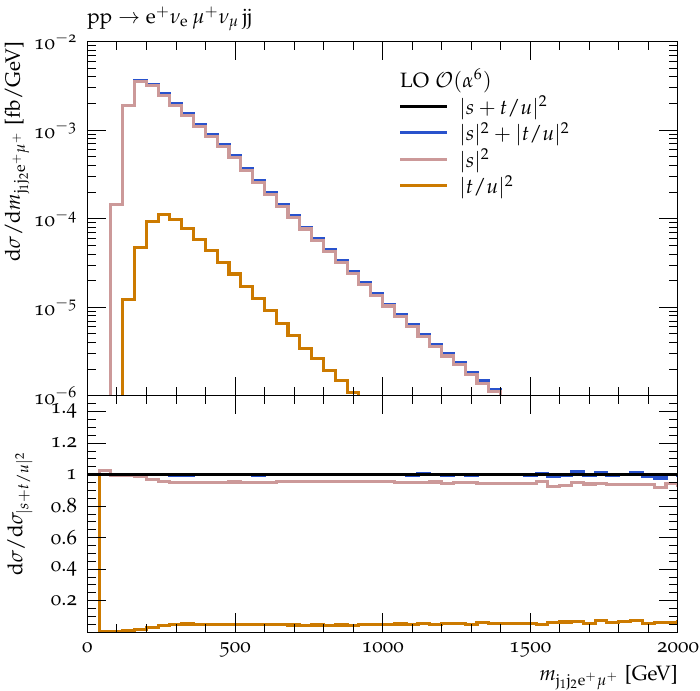}
  \end{subfigure}
  \begin{subfigure}{0.47\textwidth}
    \includegraphics[width=\textwidth]{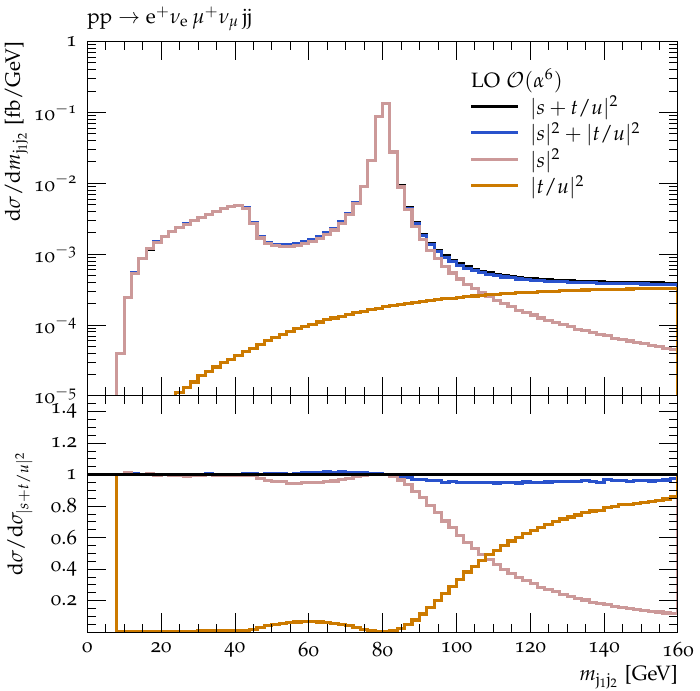}
  \end{subfigure}
  \caption{
    Differential distributions for the full off-shell process
    $\process(+\Pj)$ and its decomposition into $s$ and $t/u$~channels
    at LO $\order{\alpha^6}$, using $\muR=\muF=m_{\rT, \Pj\Pj} + m_{\rT, \nu_\Pe \Pe^+} + m_{\rT, \nu_\mu \mu^+}$.
    The observables are: 
    the
    invariant mass of the two jets and two charged leptons (left),
    and the invariant mass of the two jets (right).
    \label{fig:s-t-u-NLO-pT-m}
  }
\end{figure}

In \reffi{fig:s-t-u-NLO-pT-m} we
depict the contributions of $s$ and $t$~channels as well as their
coherent and incoherent sum for the distributions in the invariant
masses of the charged leptons and the two leading jets as well as
the di-jet system at LO $\order{\alpha^6}$.
The upper
panels show the nominal distributions and the lower ones when normalising
to the coherent sum, \ie the full off-shell
result. For the distribution in $m_{\Pj_1\Pj_2\Pe^+\mu^+}$, as for
all other distributions considered in \reffis{fig:onshell-off-shell-LO-pT-m}
and \ref{fig:onshell-off-shell-LO-dy-ct}, practically no differences
between the coherent and incoherent sum are visible. The noticeable
difference in the $m_{\Pj_1\Pj_2}$ distribution for invariant masses
above $\MW$ can be attributed to interferences between $s$-channel
and $t/u$-channel contributions.
Nonetheless, the size of the $s$--$t/u$~interference does not exceed 5\%
locally in this region, indicating the usefulness of this separation.

\subsection{Full NLO predictions}
\label{se:NLO}

We turn to NLO corrections to the full process \refeq{eq:process} which consist of four contributions at orders $\order{\alpha^7}$, $\order{\alphas\alpha^6}$, $\order{\alphas^2\alpha^5}$, and $\order{\alphas^3\alpha^4}$.
Their absolute and relative values with respect to the full LO
prediction are given in \refta{tab:NLO}.
\begin{table}
  \begin{center}
    \begin{tabular}{c||c|c|c|c||c}
     order\hl & $\order{\alpha^7}$ & $\order{\alphas\alpha^6}$ & $\order{\alphas^2\alpha^5}$ & $\order{\alphas^3\alpha^4}$ & sum \\
     \hline
     $\delta \sigma [\text{fb}]$\hl & $-0.035(1)$ & $0.305(1)$ & $-0.0032(3)$ & $0.2260(3)$ & $0.493(2)$ \\
     \hline
     $\delta \sigma / \sigma_\text{LO}^\text{sum} [\%]$\hl & $-3.4$ & $29.0$ & $-0.30$ & $21.5$ & $46.9$ \\
    \end{tabular}
  \end{center}
  \caption{\label{tab:NLO}
  Cross sections at NLO accuracy for $\process$ at the LHC for the
  four different orders and their sum using the dynamical scale of
  Eq.~\refeq{eq:scale}. The second line contains the absolute corrections
  while the third one contains the relative corrections normalised to
  the full LO prediction. 
    }
\end{table}
The largest contributions are the ones of order $\order{\alphas\alpha^6}$ and $\order{\alphas^3\alpha^4}$ and amount to $29.0\%$ and $21.5\%$, respectively.
Even if the order $\order{\alphas\alpha^6}$ is a mixture of EW and QCD corrections, it is to a good approximation dominated by the QCD ones.
Hence the NLO corrections are dominated by the QCD ones.
It is worth emphasising that the relative corrections are normalised to the full LO prediction.
If normalised to the corresponding Born contribution of order $\order{\alphas^2\alpha^4}$, the $\order{\alphas^3\alpha^4}$ corrections almost reach $90\%$.
Such large QCD corrections have actually been already observed in the literature for low invariant mass (see Figure~10 of \citere{Campanario:2013gea}).
The corrections of order $\order{\alpha^7}$ are negative and amount to 
about $3.5\%$. This is in the same ballpark as the  EW corrections for
triple-W production with fully leptonic decays
\cite{Schonherr:2018jva,Dittmaier:2019twg} or W-pair production \cite{Biedermann:2016guo}.
On the other hand, the mixed corrections at order $\order{\alphas^2\alpha^5}$ are at the per mille level.
The smallness of this contribution has also been observed in the VBS
phase space~\cite{Biedermann:2017bss} but is due to an accidental cancellation.
For VBS ZZ~\cite{Denner:2020zit,Denner:2021hsa}, these corrections are, as expected, negative at the level of few per cent with respect to the Born contribution of order $\order{\alphas^2\alpha^4}$.
In general, for the present case, the hierarchy of the corrections at full NLO accuracy is rather natural and in stark contrast to the one observed in the VBS phase space~\cite{Biedermann:2017bss}.

The full NLO prediction for the production cross section reads
\begin{align}
 \sigma_{\rm NLO} = 1.545(2)^{+6.1\%}_{-5.1\%} \; \fb.
\end{align}
The total corrections are about $47\%$, and the scale
uncertainty slightly increases with respect to the LO.
It is worth noticing that in the VBS phase space, the scale
uncertainty is $[+11.66\%$, $-9.44\%]$ at LO and $[+1.2\%,-2.7\%]$ at NLO~\cite{Biedermann:2017bss}.
In the latter case, the small NLO scale variation is related to the
small size of the QCD corrections.

At first sight, these findings seem to be in contradiction with those of \citere{Biedermann:2017bss}.
In particular, the EW corrections here have a moderate size, while they have been shown to be intrinsically large for VBS at the LHC~\cite{Biedermann:2016yds}.
In the present case, the EW corrections amount to $-4.6\%$ when normalised to the LO of order $\order{\alpha^6}$.
It is worth noticing that the photon-induced corrections contribute $+2.6\%$.
Thus, the corrections for the
$qq'$ channels are $-7.2\%$,  which is very close to the $-7.8\%$ stated in Table~3
of \citere{Dittmaier:2019twg} for the fully leptonic final state.
Also, NLO EW corrections 
for WH production have been found to be at the level of $-7\%$ \cite{Ciccolini:2003jy,Denner:2011id}. 
As explained previously, tri-boson production, WH production, and VBS share identical final states and differ only in their phase space.
In \citere{Biedermann:2016yds}, the fiducial cross section at LO is $1.5348(2)\fb$, and the relative EW corrections $-16.0\%$.
\begin{table}
\begin{center}
\begin{tabular}{l@{}l|c|c|c|c|c}
\multicolumn{3}{c|}{} & \multicolumn{2}{c|}{tri-boson phase space} &
\multicolumn{2}{c}{VBS phase space} \\
\hline 
\multicolumn{2}{c|}{partonic channel} & kin.\ & $\sigma_{\rm ch}/\sigma_{\rm all}  [\%]$ & $\delta_{\rm NLO\,EW} [\%]$ & $\sigma_{\rm ch}/\sigma_{\rm all}  [\%]$ & $\delta_{\rm NLO\,EW} [\%]$ \\[1ex]
\hline
$ \Pu   \Pu        $&${}\to \mu^+ \nu_\mu  \Pe^+ \nu_\Pe  \Pd   \Pd$ & $t,u$ & $1.2$ & $-6.3$& $67.4$ & $-17.1$ \\
$ \Pu   \Pc/\Pc   \Pu      $&${}\to \mu^+ \nu_\mu  \Pe^+ \nu_\Pe  \Pd   \Ps$  & $t$ & $0.44$ & $-5.3$ & $5.8$ & $-13.5$ \\
$ \Pu   \bar\Pd/\bar\Pd  \Pu  $&${}\to \mu^+ \nu_\mu  \Pe^+ \nu_\Pe \Pd \bar\Pu$ & $t,s$ & $49.0$ & $-7.2$ & $16.7$ & $-13.9$ \\
$ \Pu   \bar\Pd/\bar\Pd\Pu    $&${}\to \mu^+ \nu_\mu  \Pe^+ \nu_\Pe \Ps \bar\Pc $ & $s$ & $48.2$ & $-7.1$ & $0.007$ & $-30.1$ \\
$ \Pu   \bar\Ps/\bar\Ps  \Pu  $&${}\to \mu^+ \nu_\mu  \Pe^+ \nu_\Pe \Pd \bar\Pc  $ & $t$ & $0.51$ & $-5.2$ & $8.3$ & $-13.1$ \\
$ \bar\Pd  \bar\Ps/\bar\Ps  \bar\Pd $&${}\to \mu^+ \nu_\mu  \Pe^+ \nu_\Pe  \bar\Pu  \bar\Pc$ & $t$ & $0.27$ & $-3.6$ & $0.7$ & $-12.2$ \\
$ \bar\Pd  \bar\Pd $&${}\to \mu^+ \nu_\mu  \Pe^+ \nu_\Pe  \bar\Pu  \bar\Pu$ & $t,u$ & $0.31$ & $-4.1$ & $1.0$ & $-12.1$ \\
\hline
\hline
$ \Pp\Pp $&${}\to \mu^+ \nu_\mu  \Pe^+ \nu_\Pe  \Pj  \Pj$ & $-$ & $100$ & $-7.1$ & $100$ & $-16.0$ \\
\end{tabular}
\end{center}
\caption{
  Partonic channels of
  $\Pp\Pp\to\mu^+\nu_\mu\Pe^+\nu_{\Pe}\Pj\Pj$. For each of them, its
  kinematic channels are indicated. For 
  the tri-boson (present work) and the VBS phase space~\cite{Biedermann:2016yds}
  the relative contribution of each partonic channel with respect to
  the hadronic process at order $\order{\alpha^6}$ and the
  relative EW $\order{\alpha^7}$ corrections of this channel are provided.
  In this table, the partonic channels of different quark generations,
  related by simultaneously replacing $\Pu\leftrightarrow \Pc$ and $\Pd\leftrightarrow \Ps$,
  are merged together as they only differ by their PDF.
  }
\label{tab:EW-corr-by-channel}
\end{table}
In \refta{tab:EW-corr-by-channel}, absolute predictions and relative EW
corrections at order $\order{\alpha^7}$ are presented for
each partonic channel in the present setup and in the one of \citere{Biedermann:2016yds}.
It is interesting to observe that the relative EW corrections do not vary
significantly over the partonic channels.
This has already been found in \citeres{Denner:2020zit} and
\cite{Denner:2022pwc} for VBS ZZ and opposite-sign VBS WW.
Thus, large EW corrections are an intrinsic feature of VBS at the LHC, when VBS contributions are dominating.
In contrast, in phase spaces where VBS contributions are suppressed
while WH or tri-boson ones are enhanced, the EW corrections typically
are below $10\%$.

Note that the QCD corrections at order $\order{\alphas\alpha^6}$ are
significantly larger than in the VBS phase space ($29\%$ here
against $-3.5\%$ for the VBS selection).
Nonetheless, this should not come as a surprise as tenths of per cent is the usual size of QCD corrections at the LHC.
Rather, the corrections are exceptionally small for VBS.

Finally, the corrections of order $\order{\alphas^3\alpha^4}$ are much larger in the present setup than in the VBS phase space.
This can be understood from Figure~10 of \citere{Campanario:2013gea} where the QCD corrections have been found to become large when approaching the region of low invariant mass of the two jets.
In the present setup, the invariant mass of the two jets is around
the W-boson mass while in a typical VBS phase space it is required to
be above $500\GeV$. 
In \citere{Campanario:2013gea}, the reason for the large corrections
for low invariant mass is attributed to gluons splitting from quarks,
forming hence a small invariant-mass jet pair, while the other quark jet is not tagged.

\begin{figure}
\vspace{-1cm}
  \begin{subfigure}{0.5\textwidth}
    \includegraphics[width=\textwidth]{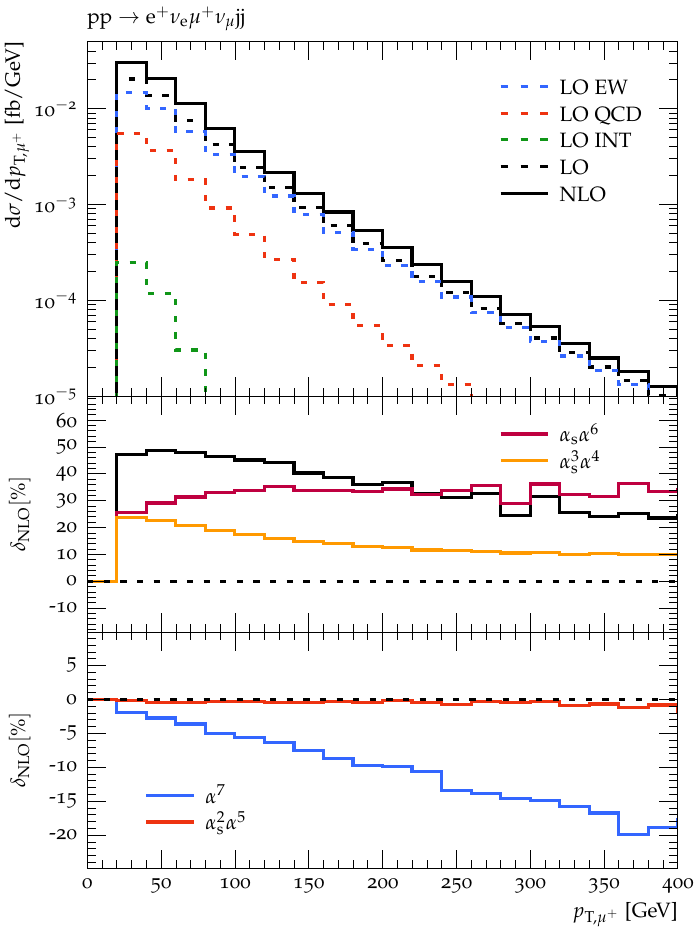}
  \end{subfigure}
  \begin{subfigure}{0.5\textwidth}
    \includegraphics[width=\textwidth]{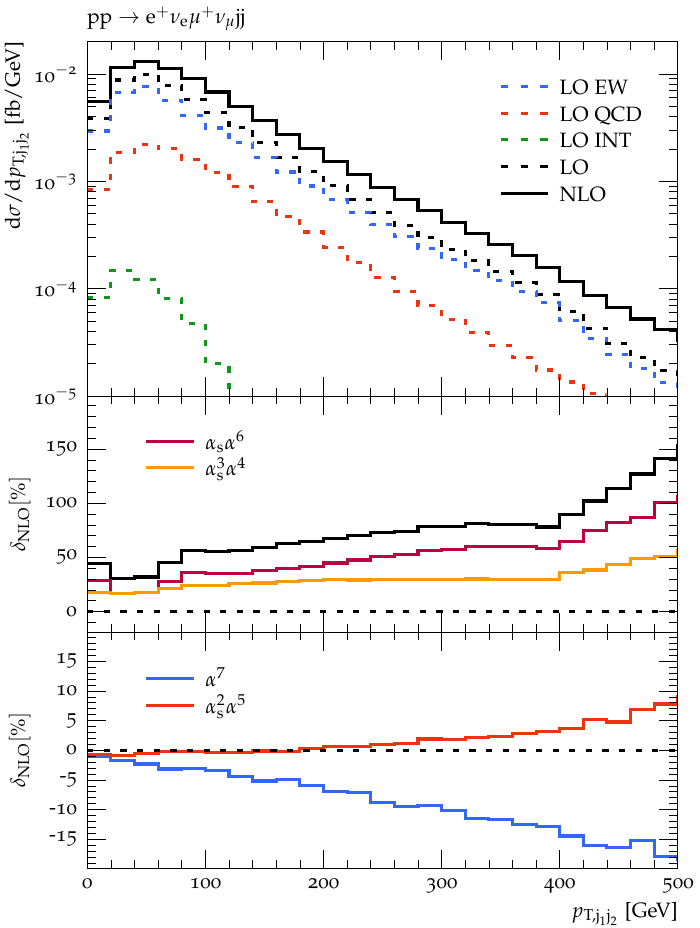}
  \end{subfigure}
  \\
  \begin{subfigure}{0.5\textwidth}
    \includegraphics[width=\textwidth]{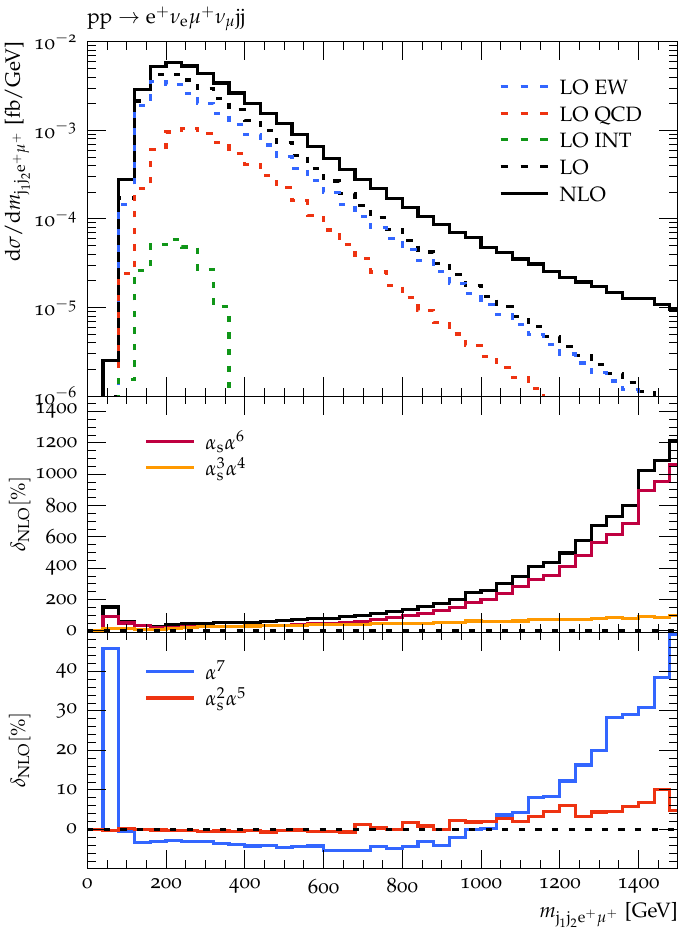}
  \end{subfigure}
  \begin{subfigure}{0.5\textwidth}
    \includegraphics[width=\textwidth]{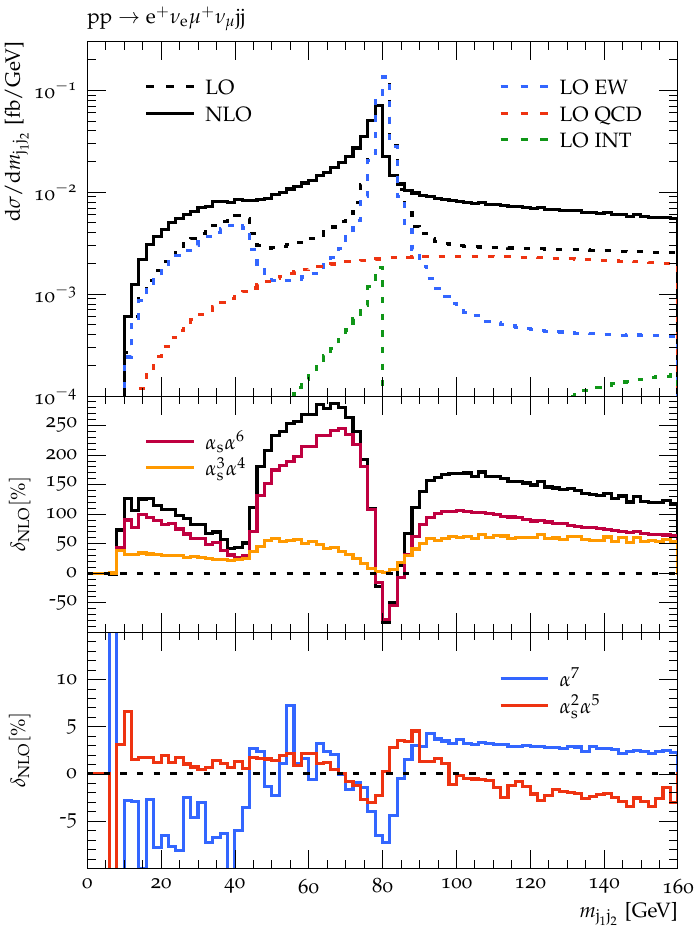}
  \end{subfigure}
  \caption{\label{fig:full_nlo_separated}%
  Differential distributions at full NLO accuracy for $\process$.
  The observables are: the transverse momentum of the anti-muon (top
  left), the transverse momentum of the two jets (top right), the
  invariant mass of the two jets and two charged leptons (bottom
  left), and the invariant mass of the two jets (bottom right).
  }
\end{figure}
\begin{figure}
  \begin{subfigure}{0.5\textwidth}
    \includegraphics[width=\textwidth]{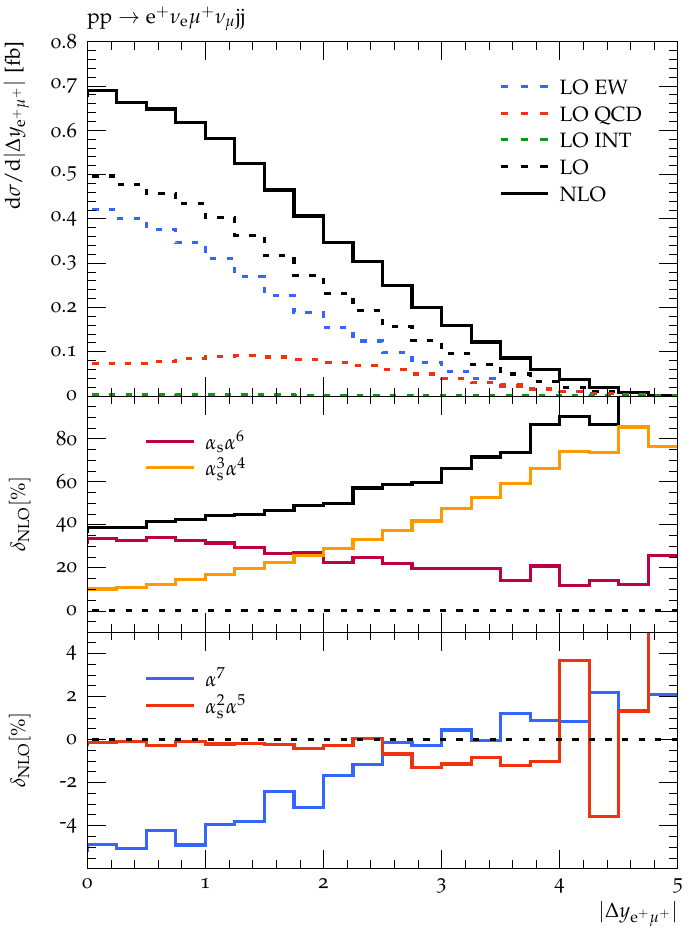}
  \end{subfigure}
  \begin{subfigure}{0.5\textwidth}
    \includegraphics[width=\textwidth]{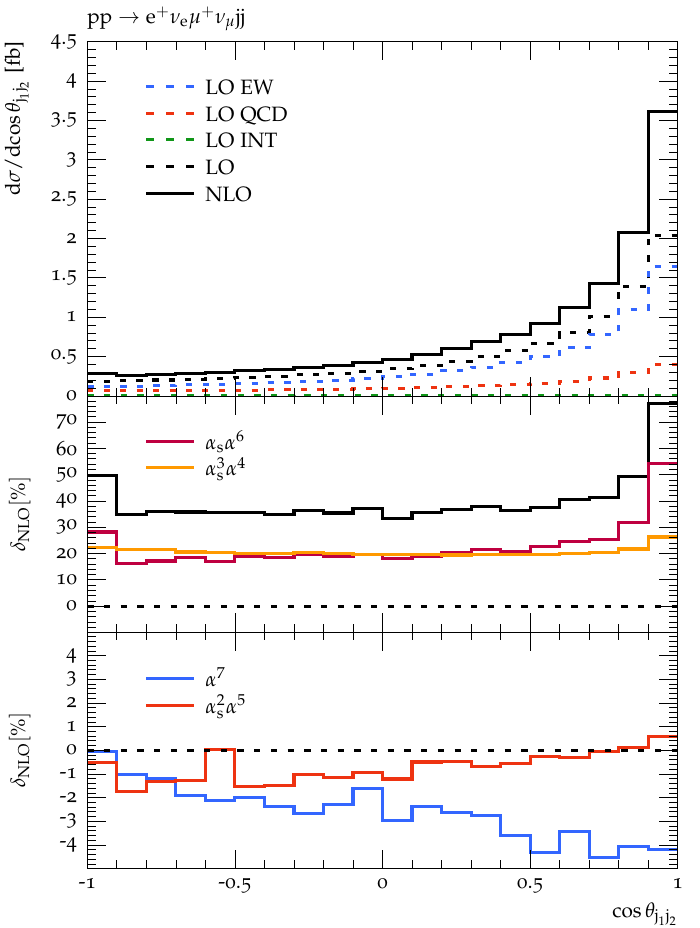}
  \end{subfigure}
  \caption{\label{fig:full_nlo_separated_2}%
  Differential distributions at full NLO accuracy for $\process$.
  The observables are: the modulus of the rapidity difference between the two charged leptons (left) and the cosine of the angle between the two jets (right).
  }
\end{figure}

In the following, several differential distributions are discussed.
First, in \reffis{fig:full_nlo_separated} and \ref{fig:full_nlo_separated_2},
all LO and NLO contributions are shown separately.
While the upper panels contain the full NLO predictions and the
separate LO contributions, the middle and lower panels show the full NLO
corrections relative to the full LO along with the contributions of
individual NLO orders.

The distribution in the transverse momentum of the anti-muon
(top~left in \reffi{fig:full_nlo_separated}) rapidly falls off towards high energy.
The $\order{\alphas\alpha^6}$ corrections exceed $20\%$ in the lowest bin and increase up to almost $40\%$ at $200\GeV$ where they reach a plateau.
The $\order{\alphas^3\alpha^4}$ corrections are at the same level as the $\order{\alphas\alpha^6}$ ones in the first bin but decrease for higher transverse momenta.
The $\order{\alphas^2\alpha^5}$ corrections are essentially zero across the whole phase space as for the fiducial cross section.
The $\order{\alpha^7}$ corrections, on the other hand, display the typical Sudakov behaviour with negative corrections reaching $-15\%$ at $p_{\rT,\mu^+}\approx 300\GeV$.

For the distribution in the transverse momentum of the two jets (top~right in \reffi{fig:full_nlo_separated}), the $\order{\alpha^7}$ corrections display a similar qualitative behaviour.
On the other hand, the $\order{\alphas^2\alpha^5}$ corrections are at the
per-mille level up to $250\GeV$ and thereafter increase slightly to reach $5\%$ close to $450\GeV$ and almost $15\%$ for $\ptsub{\Pj_1\Pj_2}=600\GeV$ (not shown).
The corrections of order $\order{\alphas^3\alpha^4}$ increase to
almost $30\%$ at $400\GeV$, while the $\order{\alphas\alpha^6}$ ones
are larger, reaching about $60\%$ at this transverse momentum.
At about $70\GeV$ in the second and third bin, there is a dip in the distribution, which is driven uniquely by the corrections of order $\order{\alphas\alpha^6}$.
This is related to the presence of the W boson which decays
hadronically, a configuration which does not exist for the QCD-induced
contribution. While at LO the two leading jets dominantly result from
the decay of the W~boson that originates from a Higgs boson, at NLO one of the
leading jets can be a bremsstrahlung jet and $\ptsub{\Pj_1\Pj_2}$ does
not correspond to the transverse momentum of the W~boson and tends to
be higher. The stronger increase of the relative corrections above
$400\GeV$ is related to a faster drop of the LO EW cross section in this region.
The latter results from an interplay of the cut \refeq{eq:jet2}
on $m_{\Pj\Pj}$ and the jet recombination parameter $R=0.4$. For
$\ptsub{\Pj\Pj}\gsim400\GeV$, jet pairs with $m_{\Pj\Pj}<160\GeV$ get more
and more recombined and the corresponding events are cut away.

The distribution in the invariant mass of the visible system (two charged leptons and two jets, shown in the lower-left panel of \reffi{fig:full_nlo_separated}) is particularly interesting.
At high energy, the corrections of orders $\order{\alphas^3\alpha^4}$ and $\order{\alphas\alpha^6}$ become very large.
This is particularly true for the latter one which exceeds $100\%$ at $900\GeV$.
This dramatic effect is due to contributions of $t$-channel
topologies, \ie partonic processes where quark lines run from the initial
to the final state (see \reffis{diag:real_gluon_from_quark_VV} and \ref{diag:real_gluon_from_quark_VV2}). The hadronically decaying W~boson is faked by a
quark-gluon pair, while the second quark jet is subleading or cut
away. These diagrams are enhanced by a $t$-channel W~boson similarly
to VBS topologies (see \reffi{diag:real_gluon_from_quark_VBS}). Obviously, such contributions
are neither present in genuine triple-W-production contributions nor
at LO in the full process.
It is worth noticing that the EW corrections of order $\order{\alpha^7}$ do not become negatively large under the influence of Sudakov logarithms in the high-energy region of this distribution.
The reason for this is twofold: On the one hand, photon-induced contributions become large,
reaching $+20\%$ at $1\TeV$, and compensate the negative Sudakov-like contributions.
Note that at this energy, the cross section is strongly reduced, and
so even if relatively significant, photon-induced contributions remain
small in absolute terms.
On the other hand, a large invariant mass does not imply the
Sudakov regime, because some invariants may still be rather small
\cite{Biedermann:2016guo}. It has been demonstrated that
logarithmic terms of the form $\alpha\log^2(t/s)$ may cause large
corrections for processes involving $t$-channel propagators
\cite{Pagani:2021vyk,Lindert:2023fcu} that can
be of the order of $10\%$ at LHC energies \cite{Lindert:2023fcu}.

The distribution in the invariant mass of the two leading jets
(bottom~right in \reffi{fig:full_nlo_separated}) receives large corrections below the W-mass peak.
This is characteristic for final-state radiation that carries away
energy and therefore shifts events from the peak to below it.
In addition, large corrections of order $\order{\alphas\alpha^6}$ [and to a lesser extend $\order{\alphas^3\alpha^4}$] are observed above the peak.
This region opens up at NLO, owing to the decaying quarks of the W boson recombining into one jet while the QCD radiation makes up the second hard jet \cite{Ballestrero:2018anz,Denner:2020zit}, while it is suppressed at LO.
The EW corrections of $\order{\alpha^7}$ have a similar structure as
the corrections of order $\order{\alphas\alpha^6}$ albeit at a reduced
level but do not rise towards very small jet--jet invariant masses.
The even smaller variation of the relative $\order{\alphas^2\alpha^5}$
corrections follows the one of $\order{\alphas^3\alpha^4}$ except for
the region of high $m_{\Pj_1\Pj_2}$.
As mentioned above (see \reffi{fig:onshell-off-shell-LO-pT-m}), the
peak in the distribution around $40\GeV$ is due to the $\PW\PH$ contribution \cite{Bredenstein:2006ha,Denner:2022pwc}.
All in all, we can observe that the full LO structure with two sharp structures is strongly distorted by higher-order corrections which smear these shapes.

For the distribution in the rapidity difference of the two charged leptons, presented in
the left panel of \reffi{fig:full_nlo_separated_2},
the $\order{\alphas^2\alpha^5}$ corrections remain very small across the whole phase space.
The EW corrections of order $\order{\alpha^7}$ are only negative in the region of low rapidity difference where the bulk of the cross section sits.
This is due to the large positive  ($\sim +7\%$) photon-induced contribution in the high-rapidity regions.
Finally, the corrections of order $\order{\alphas^3\alpha^4}$ and
$\order{\alphas\alpha^6}$ display an opposite behaviour.
The former are minimal at low rapidity differences and larger at high
differences, while the latter reach their maximum for small and
decrease for larger rapidity separations.

Finally, the distribution  in the cosine of the angle between the two jets
(right in \reffi{fig:full_nlo_separated_2})
is maximal at $\cos \theta_{\Pj\Pj} \sim 1$, \ie when the two jets are close to each other.
All the different NLO corrections show only small variations in this distribution with the noticeable exception of those of order $\order{\alphas\alpha^6}$, which increase towards $\cos \theta_{\Pj\Pj} \sim 1$ and reach about $55\%$ in the last bin.
This results again from real contributions with a gluon jet collinear to a quark
jet that fake the hadronically decaying W~boson and thus evade the
invariant-mass cut \refeq{eq:jet2}.
Note that in this region, the EW corrections are compensated by photon-induced contributions, which reach $+4\%$ in the right-most bin.

\begin{figure}
  \begin{subfigure}{0.5\textwidth}
    \includegraphics[width=\textwidth]{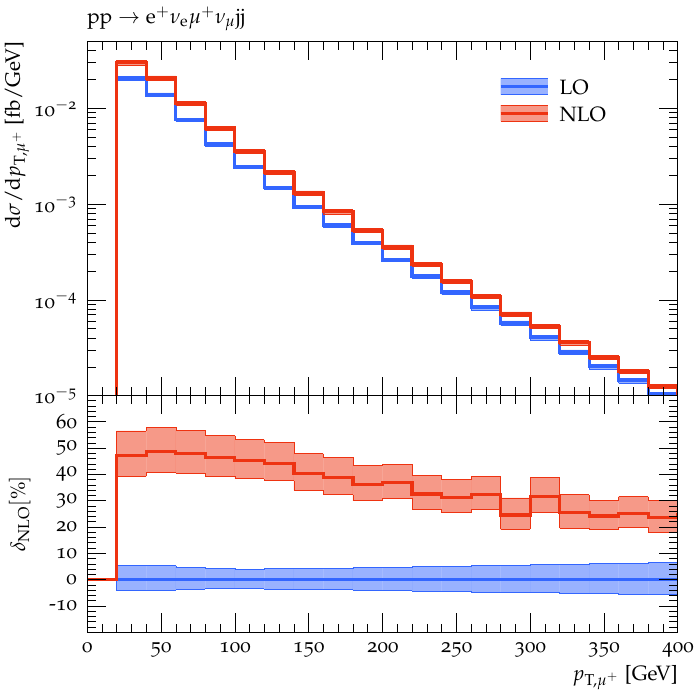}
  \end{subfigure}
  \begin{subfigure}{0.5\textwidth}
    \includegraphics[width=\textwidth]{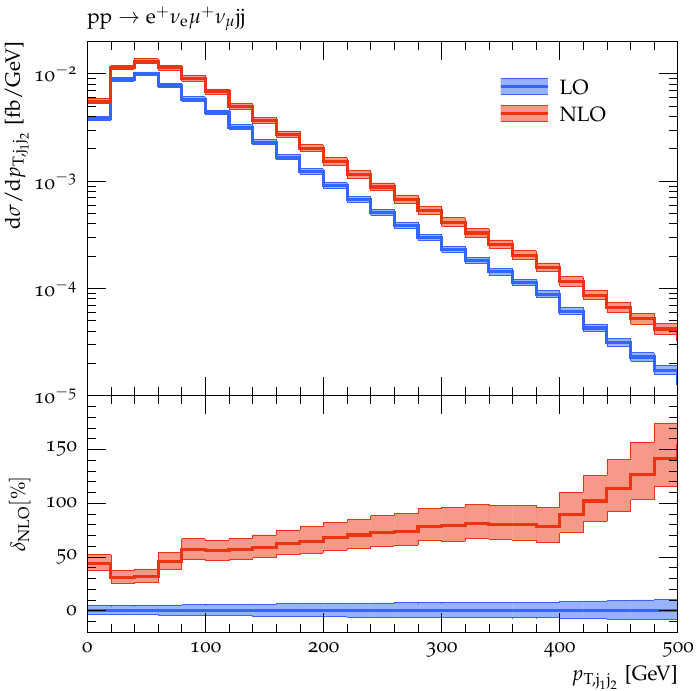}
  \end{subfigure}
  \\
  \begin{subfigure}{0.5\textwidth}
    \includegraphics[width=\textwidth]{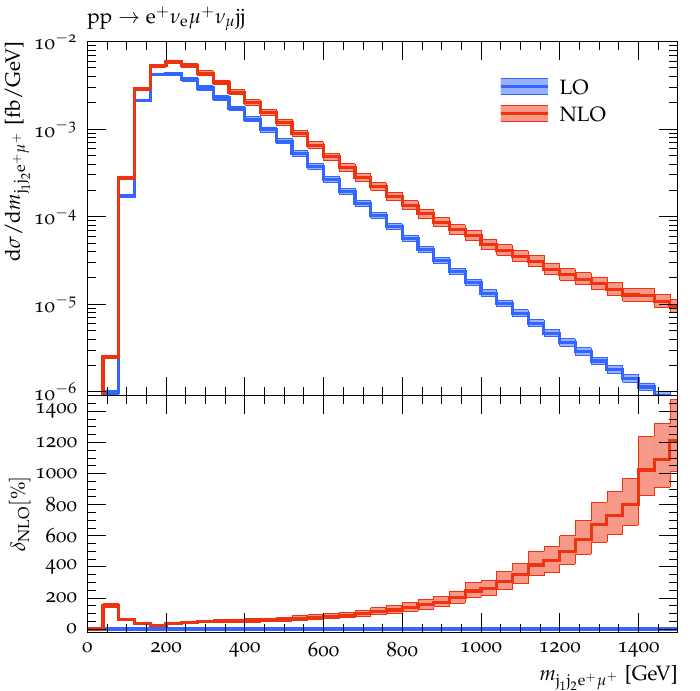}
  \end{subfigure}
  \begin{subfigure}{0.5\textwidth}
    \includegraphics[width=\textwidth]{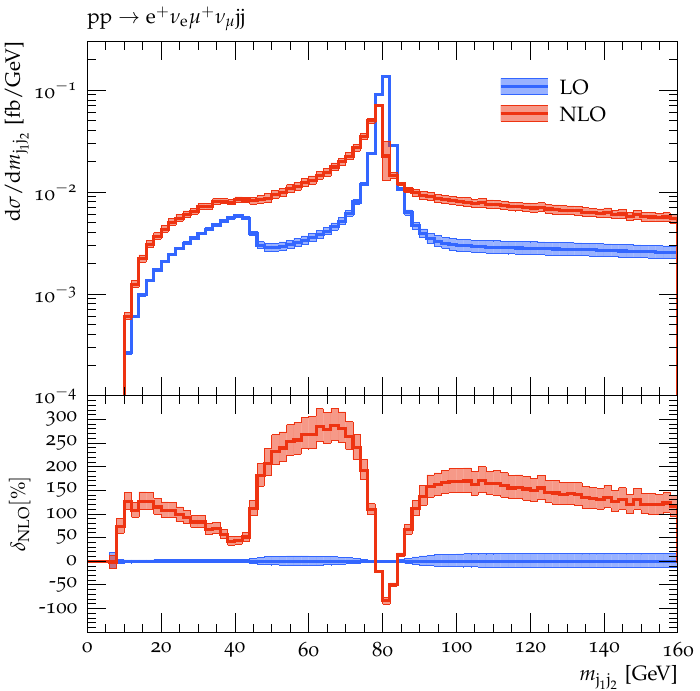}
  \end{subfigure}
        \caption{\label{fig:full_nlo_all}%
  Differential distributions at full NLO accuracy (combined) for $\process$.
  The observables are: the transverse momentum of the anti-muon (top
  left), the transverse momentum of the two jets (top right), the
  invariant mass of the two jets and two charged leptons (bottom
  left), and the invariant mass of the two jets (bottom right).}
\end{figure}%
\begin{figure}
  \begin{subfigure}{0.5\textwidth}
    \includegraphics[width=\textwidth]{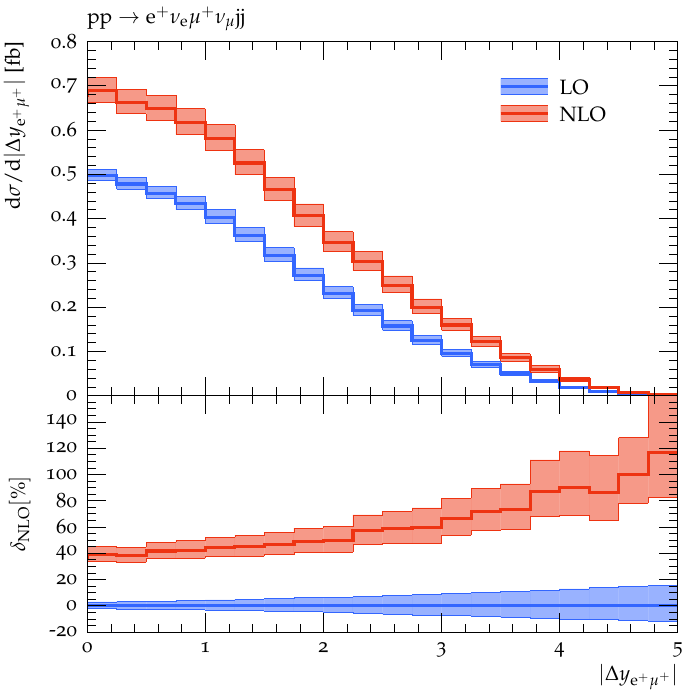}
  \end{subfigure}
  \begin{subfigure}{0.5\textwidth}
    \includegraphics[width=\textwidth]{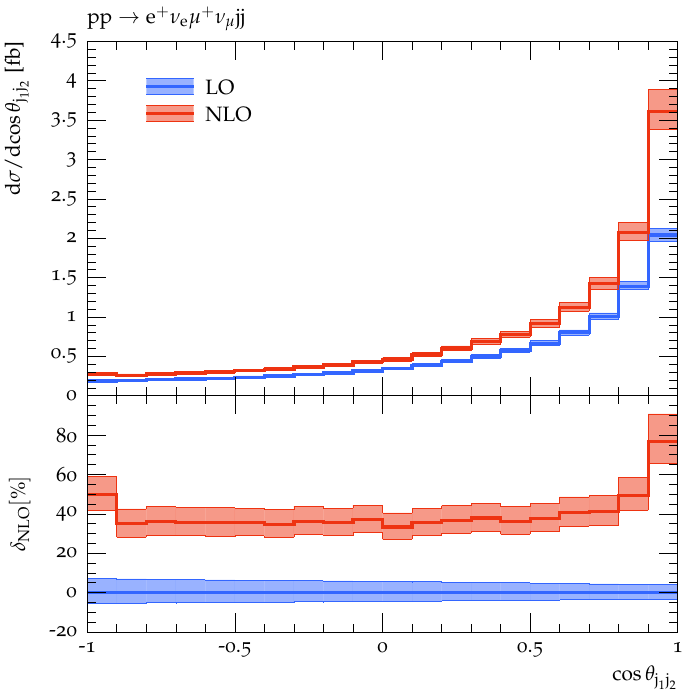}
  \end{subfigure}
  \caption{\label{fig:full_nlo_all_2}%
     Differential distributions at full NLO accuracy (combined) for $\process$.
      The observables are: the modulus of the rapidity difference between the two charged leptons (left) and the cosine of the angle between the two jets (right).}
\end{figure}%
In \reffis{fig:full_nlo_all} and \ref{fig:full_nlo_all_2}, the same distributions as in
\reffis{fig:full_nlo_separated} and \ref{fig:full_nlo_separated_2}
are displayed, but for the full NLO and 
LO predictions including the respective scale uncertainty. While the
upper panels show the absolute LO and NLO predictions, the lower
panels display these contributions normalised to the LO predictions at
the central scale.
For the distribution in the transverse momentum of the anti-muon
(top~left of \reffi{fig:full_nlo_all}),
the NLO corrections are around $50\%$ at low values and decrease smoothly under the influence of the $\order{\alphas^3\alpha^4}$ and $\order{\alpha^7}$ corrections to reach $20\%$ at $400\GeV$.
The NLO corrections to the distribution in the transverse momentum of
the two jets (top~right of \reffi{fig:full_nlo_all}) display an opposite behaviour.
At about $50\GeV$, there appears the dip originating from the
$\order{\alphas\alpha^6}$ corrections discussed above. For smaller $\ptsub{\Pj\Pj}$, the total corrections are slightly below $50\%$ while above they grow very large to exceed $80\%$ above $400\GeV$.
For the distribution in the invariant mass of the visible system (bottom~left of \reffi{fig:full_nlo_all}), the corrections are also very large.
They are at the level of $80\%$ at $600\GeV$, and the NLO cross section is an order of magnitude larger than the LO one at $1.5\TeV$.
The distribution in the invariant mass of the two jets (bottom~right of
\reffi{fig:full_nlo_all}) receives
positive corrections apart from the bin around the W~mass where the
corrections are at the level of $-50\%$.
This feature is largely due to the $\order{\alphas\alpha^6}$ corrections.
The corrections to the distribution in the rapidity difference of the
two charged leptons (left of \reffi{fig:full_nlo_all_2}) are significant.
They are minimal for small $|\Delta y_{\Pe^+\mu^+}|$, where they reach $40\%$ in the bulk of the cross section.
For more extreme phase-space regions with large rapidity differences, the corrections are very large and grow up to almost $100\%$.
Finally, the corrections to the distribution in the cosine of the angle between the two jets  (right of \reffi{fig:full_nlo_all_2}) are rather flat in most of the phase space.
They are at the level of $40\%$ almost everywhere apart from the two
bins close to $\cos \theta = 1$, where most of the cross section is
located and where the corrections reach $80\%$ mainly driven by the
$\order{\alphas\alpha^6}$ corrections. Note that the NLO
corrections are in general much larger than the LO scale variation,
\ie the scale variation does not provide a good  measure of the
uncertainty of the calculation.

\subsection{NLO QCD matched to parton shower with virtual \EW\ approximation}
\label{se:mcatnlo}

In this section, parton-shower-matched predictions for
off-shell $\process$ production, supplemented by EW corrections in the
virtual \EW\ approximation, are presented.
In particular, they are compared against the full fixed-order results.
While in a full computation,  EW and QCD contributions cannot be
separated unambiguously beyond LO \cite{Biedermann:2017bss}, this
distinction is nonetheless typically made in experimental analyses
or new-physics studies.
To this end, we follow the approach of \citere{Lindert:2022ejn}
and separate all contributions (see \reffi{fig:allorders}) into
QCD and EW production processes as follows:
\begin{description}
  \item[QCD production.] We define the QCD production mode at
    LO by the $\order{\alpha_s^2\alpha^4}$ process.
    Its QCD corrections include all NLO contributions of
    $\order{\alpha_s^3\alpha^4}$ while those of
    $\order{\alpha_s^2\alpha^5}$ constitute its EW corrections.
    The latter can also be understood as QCD
    corrections to the LO QCD--EW interference of $\order{\alpha_s\alpha^5}$
    and, accordingly, contain diagrams with
    up to three resonant \PW~bosons interfered with diagrams involving
    a gluon exchange and at most two resonant \PW~bosons, \eg the interference
    of diagrams illustrated in \reffis{diag:loopw:gs2} and
    \ref{diag:LO:QCDs}.
    Our parton-shower-matched \MCatNLO calculation includes the
    full off-shell LO and NLO QCD contributions, but the
    $\order{\alpha_s^2\alpha^5}$ EW corrections, owing to this feature,
    are omitted. However, as shown in \refse{se:NLO}, the
    $\order{\alpha_s^2\alpha^5}$ corrections are rather small.
    Nonetheless, QED corrections are accounted for through
    \Sherpa's YFS soft-photon resummation. Note, in the \MCatNLO
    simulation of this QCD production process, in contrast to the fixed-order
    calculations, all QCD couplings get evaluated at the reconstructed emission
    scale associated to the corresponding parton, \ie its relative transverse
    momentum~\cite{Hoeche:2009rj,Hoche:2010kg}. This applies to the two
    powers of $\alpha_s$ present at Born level, the QCD NLO correction,
    as well as all subsequent parton-shower splittings.
  \item[EW production.] Likewise, the EW production mode is defined
    at LO by the $\order{\alpha^6}$ process, and its QCD and EW
    corrections at $\order{\alpha_s\alpha^6}$ and $\order{\alpha^7}$,
    respectively.
    This time, the QCD corrections can also be interpreted as the EW
    corrections to the LO QCD--EW interference, containing, for example 
    interferences of diagrams with three resonant $s$-channel \PW~bosons
    (such as illustrated in \reffi{diag:LO:WWW}) and those missing the
    hadronic resonant decay (see \reffi{diag:loopz:gs2}).
    The parton-shower-matched \MCatNLO calculation is split into its
    $s$- and $t/u$-channel components, as explained in \refse{se:s-t-channels},
    but otherwise treated fully off shell.
    We neglect the  $s$--$t/u$-channel interference, which has been found to be small in \refse{se:s-t-channels}.
    Foremost, this separation implies that QCD radiation
    off the $s$-channel resonant \PW decay does not interfere with
    any other source of radiation, allowing for a standard matching
    prescription as the virtuality of intermediate resonances is preserved.
    Likewise, it removes all EW-type divergences
    described above in the $\order{\alpha_s\alpha^6}$ corrections,
    reducing the complexity of the combination with the parton shower
    to a standard QCD \MCatNLO matching.
    In each case, the calculation contains the full respective LO and
    NLO QCD contributions, while the EW corrections are added in the
    \EWvirt\ scheme. As above, \Sherpa's YFS soft-photon resummation
    provides QED corrections. As for the QCD production channel, in the
    matched calculation all coupling factors for QCD emissions get evaluated
    at their relative transverse momentum. For the EW production mode, this
    affects the NLO QCD correction and all parton-shower emissions. 
\end{description}
In line with the above findings, we neglect
the LO QCD--EW interference of $\order{\alpha_s\alpha^5}$ in the
parton-shower-matched predictions.

\begin{table}
  \begin{center}
    \begin{tabular}{r|c|c||C|D||C|E}
      \multicolumn{3}{c||}{} &
      \multicolumn{2}{c||}{NLO} &
      \multicolumn{2}{c}{\MCatNLO} \\\cline{4-7}
      \multicolumn{3}{c||}{}
      & QCD & QCD+EW & QCD & QCD+\EWvirt \hl \\
      \hline\hline
      \parbox[t]{2mm}{\rotatebox[origin=c]{90}{\,QCD prod.\,}}
      & $\sigma [\text{fb}]$ &
      & $0.485$ & $0.482$ & $0.484$ & -- \\
      \hline\hline
      \parbox[t]{2mm}{\multirow{4}{*}{\rotatebox[origin=c]{90}{EW prod.}}}
      & \multirow{4}{*}{$\sigma [\text{fb}]$}
      & $|s+t/u|^2$\hl
      & $1.091$ & $1.056$ & \multicolumn{2}{c}{}  \\\cline{3-7}
      &
      & $|s|^2+|t/u|^2$\hl
      & $1.084$ &  & $1.017$ & $0.973$ \\\cline{3-4}\cline{6-7}
      &
      & $|s|^2$\hl
      & $0.998$ & -- & $0.882$ & $0.842$ \\\cline{3-4}\cline{6-7}
      &
      & $|t/u|^2$\hl
      & $0.086$ &  & $0.135$ & $0.131$
    \end{tabular}
  \end{center}
  \caption{
    Comparison of fiducial cross sections for the QCD and EW production processes at
    full NLO and from \MCatNLO calculations with \Sherpa. For the EW production \MCatNLO
    result a break down into the kinematic $s$- and $t/u$-channel contributions is
    provided. \label{tab:MCNLO}
  }
\end{table}

In \refta{tab:MCNLO} we collate fiducial cross sections for the
QCD and EW production modes, comparing the results obtained with the
shower-matched calculation and the off-shell NLO QCD and full NLO
prediction.
Considering the separation into QCD and EW production, we
here ignore the mixed LO contribution of order $\order{\alpha_s\alpha^5}$,
which contributes about $0.007\,\text{fb}$ to the total cross section
(see~\refta{tab:LO}).
Notably, for the QCD production mode the full NLO QCD calculation and
the \MCatNLO result agree very well to within 1\%.
We do not include EW corrections for the shower-matched calculations
for this component, which, however, for the full calculation amount
to $-0.6\%$ only.

In the EW production process, at NLO QCD accuracy, the two calculations
differ by $7\%$ ($1.017\fb$ versus $1.091\fb$), 
with the \MCatNLO simulation predicting a lower cross
section.
When studying the quality of the incoherent $s$- and $t/u$-channel
approximation in \refse{se:s-t-channels} for the fixed-order
calculation, we instead observed a difference between the coherent
and the incoherent result of only $0.6\%$ (see~\refta{tab:NLO-s-t-channel}).
In consequence, the larger difference observed is attributed to
the additional parton-shower corrections beyond NLO accuracy.
Similarly, the admixture of the $s$- and $t/u$-channel contributions is
altered in the \MCatNLO simulation.
While the $t/u$-channel component contributes about $8\%$ to the
fixed-order NLO result, it makes up $13\%$ of the full
\MCatNLO result.
This is a consequence of the fact that the $t/u$-channel process
already experienced much larger NLO corrections than the $s$-channel
process which was driven by the emission of an additional parton as
that parton made the presence of a central jet-pair within the
required mass window much more likely.
The scale choice for the QCD coupling in the \MCatNLO calculation and
additional shower emissions further increase this probability and hence the
cross section of this subprocess by another 52\% wrt.\ the fixed-order
NLO QCD result.
Conversely, the $s$-channel process loses 12\% of its events
through the explicit modelling of
additional radiation with the parton shower.
Explicitly resolving the multiple-emission kinematics leads
to a more precise modelling of the radiative energy loss of the
jet through out-of-cone radiation, \ie radiation at angles large enough
not to be recaptured by the jet recombination procedure,
leading to a significant
number of jets falling below the jet-$p_\mathrm{T}$ threshold
and thus reducing the fiducial cross section.
Through their respective characteristic di-jet correlations
this leads to a larger impact in the $s$~channel, dominated by
$\PW\PW\PW$ and $\PW\PH$ topologies, as compared to the $t$~channel.
Taken together, this changes not only the composition of the combined
sample, but also reduces its combined cross section by about 6\% 
($1.017\fb$ versus $1.084\fb$).

The EW corrections to the EW production process are $-3.2\%$ at
fixed order, of which $+1.9\%$ are contributed by the photon-induced
channels, while they amount to $-4.3\%$ for the full \MCatNLO
result, where the photon-induced channels are not accounted for.
For the $s$-channel \MCatNLO we obtain $-4.5\%$ and for the
$t$-channel contribution $-3.0\%$.
In light of the fact that the \EWvirt\ approximation is designed to
recover the NLO EW corrections in the Sudakov limit, this agreement
for the inclusive cross section is reasonable and does not spoil
the overall accuracy of the predictions.
Altogether, the final \Sherpa prediction of NLO QCD+\EWvirt\ accuracy
is $8\%$ lower than the NLO QCD+EW result. 

\begin{figure}
  \centering
  \begin{subfigure}{0.47\textwidth}
    \includegraphics[width=\textwidth]{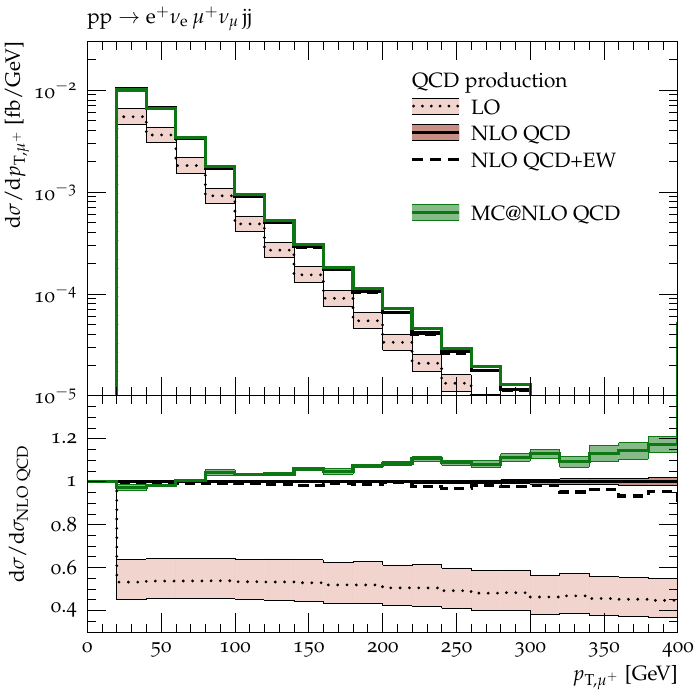}
  \end{subfigure}
  \begin{subfigure}{0.47\textwidth}
    \includegraphics[width=\textwidth]{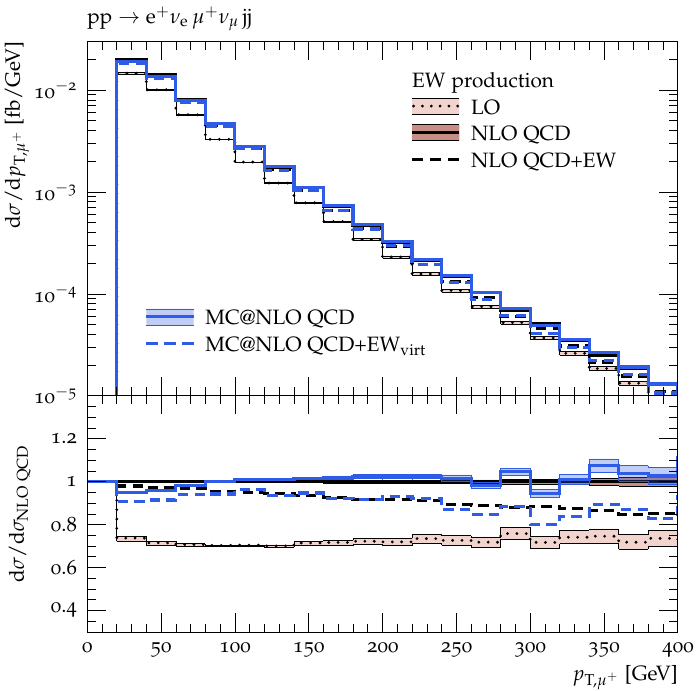}
  \end{subfigure}
  \\
  \begin{subfigure}{0.47\textwidth}
    \includegraphics[width=\textwidth]{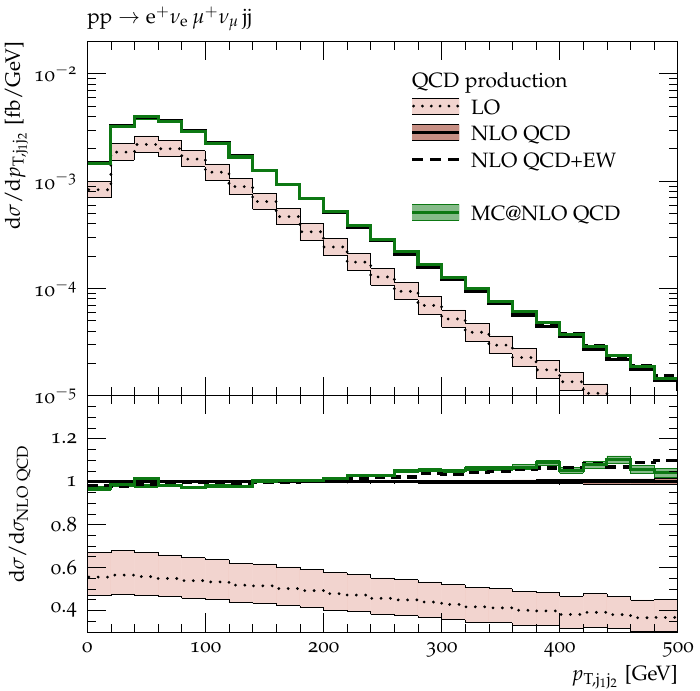}
  \end{subfigure}
  \begin{subfigure}{0.47\textwidth}
    \includegraphics[width=\textwidth]{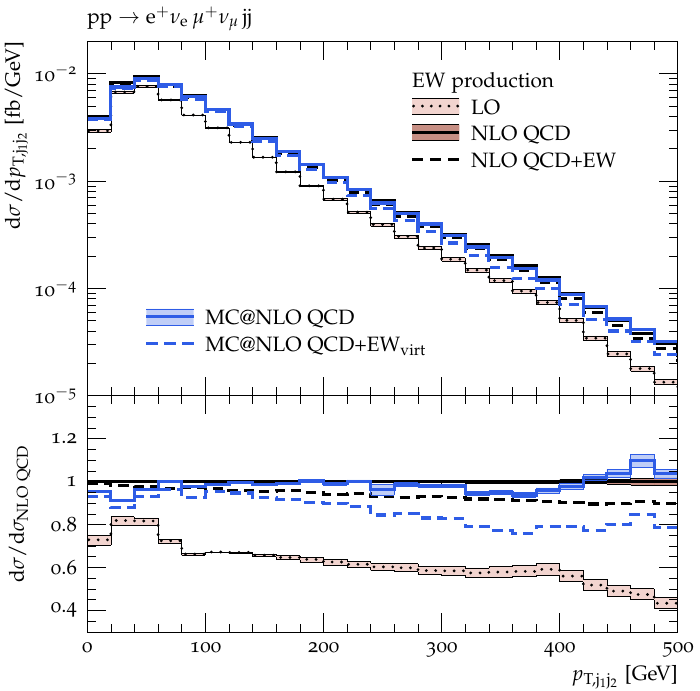}
  \end{subfigure}
  \caption{
    Parton-shower-matched predictions for the QCD and EW production
    modes contributing to $\process$. For the QCD channel (left panels)
    we compare the \MCatNLO results at NLO QCD accuracy obtained from
    \Sherpa with the LO and NLO QCD and QCD+EW predictions. For the
    EW channel (right panels) we in addition include approximate EW
    corrections in the \MCatNLO calculation, labelled as
    \MCatNLO QCD+\EWvirt. Results are shown for the transverse momentum
    of the anti-muon (top row) and the transverse momentum of the di-jet
    system (bottom row). 
    \label{fig:mcatnlo-pT}
  }
\end{figure}
\begin{figure}
  \centering
  \begin{subfigure}{0.47\textwidth}
    \includegraphics[width=\textwidth]{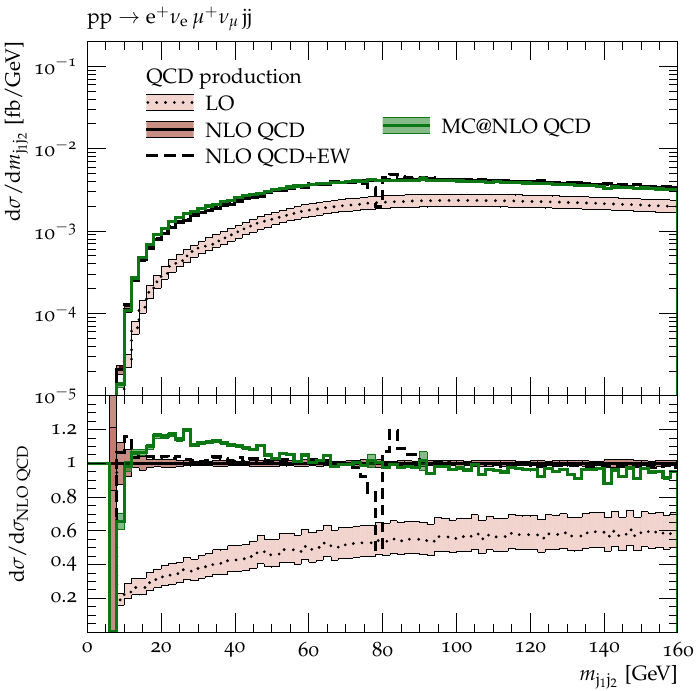}
  \end{subfigure}
  \begin{subfigure}{0.47\textwidth}
    \includegraphics[width=\textwidth]{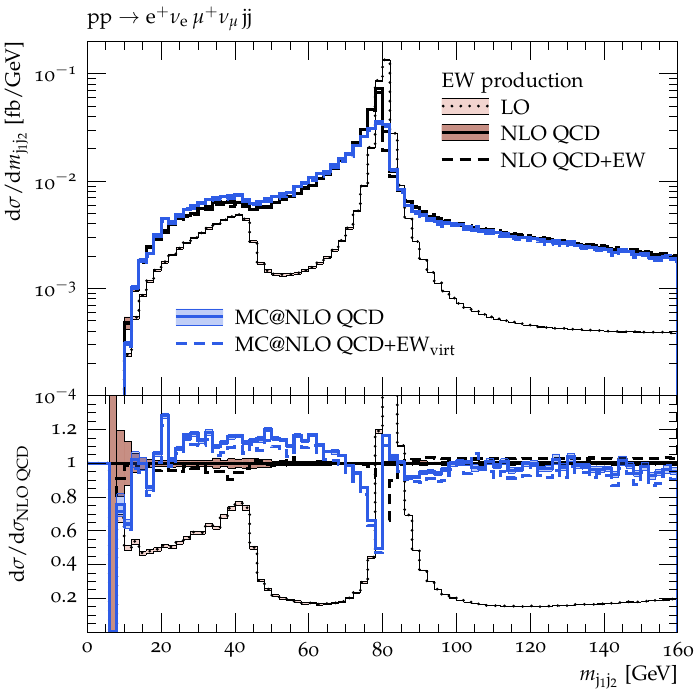}
  \end{subfigure}
  \\
  \begin{subfigure}{0.47\textwidth}
    \includegraphics[width=\textwidth]{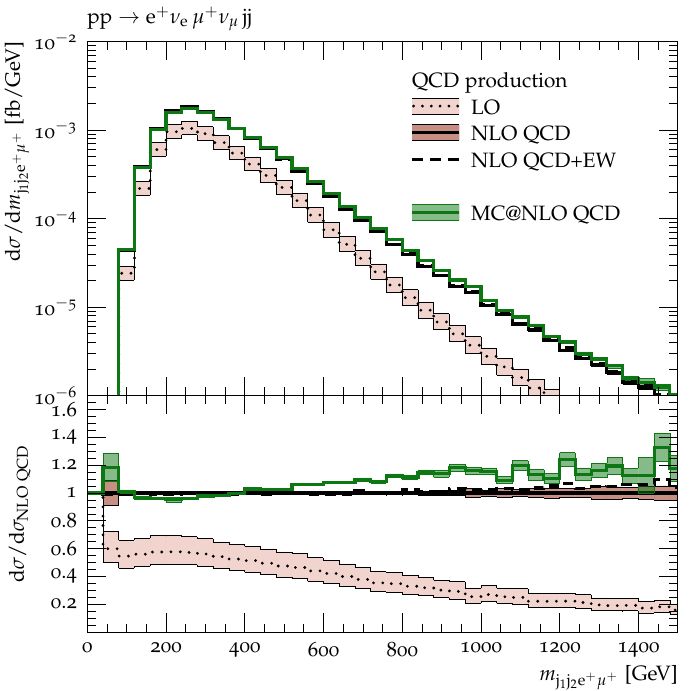}
  \end{subfigure}
  \begin{subfigure}{0.47\textwidth}
    \includegraphics[width=\textwidth]{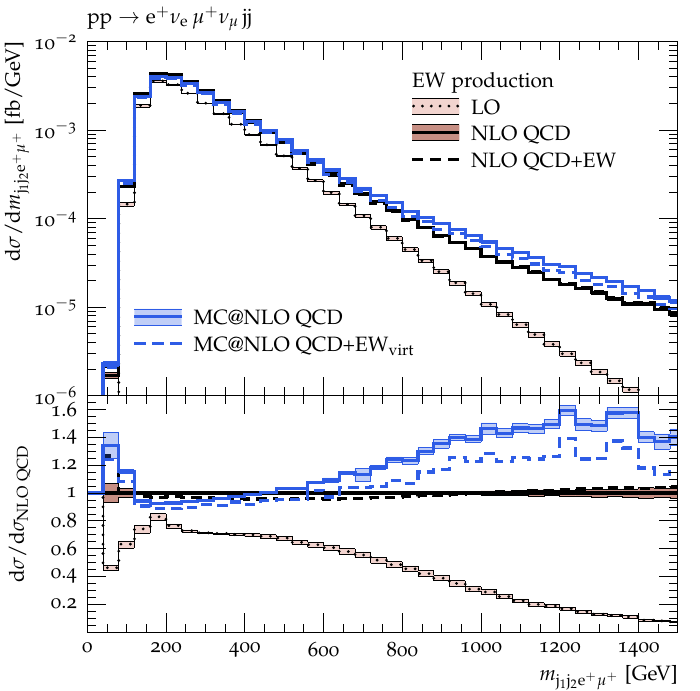}
  \end{subfigure}
  \caption{
    Parton-shower-matched predictions for the QCD and EW production
    modes contributing to $\process$ in comparison to LO and NLO QCD
    and NLO QCD+EW predictions. Results are shown for the invariant
    masses of the di-jet system (top row) and the system formed by the
    two leading jets and the two charged leptons (bottom row). 
    \label{fig:mcatnlo-m}
  }
\end{figure}
\begin{figure}
  \centering
  \begin{subfigure}{0.47\textwidth}
    \includegraphics[width=\textwidth]{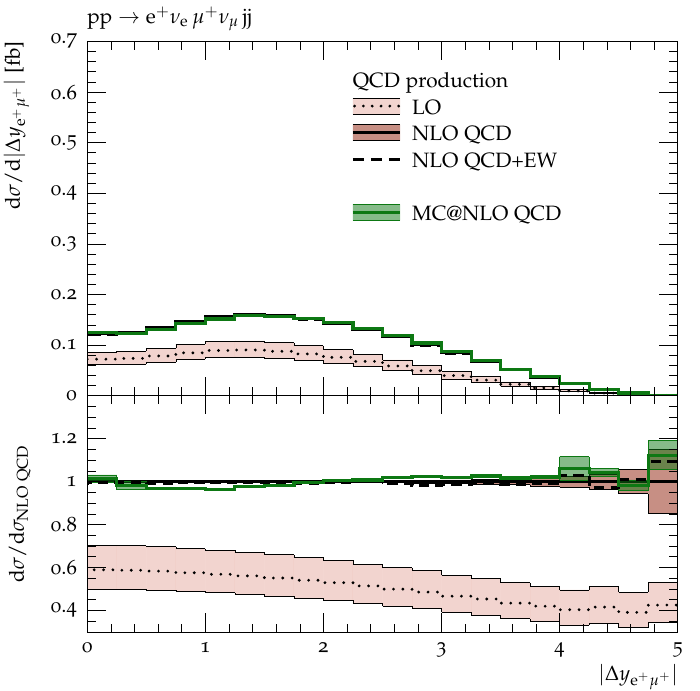}
  \end{subfigure}
  \begin{subfigure}{0.47\textwidth}
    \includegraphics[width=\textwidth]{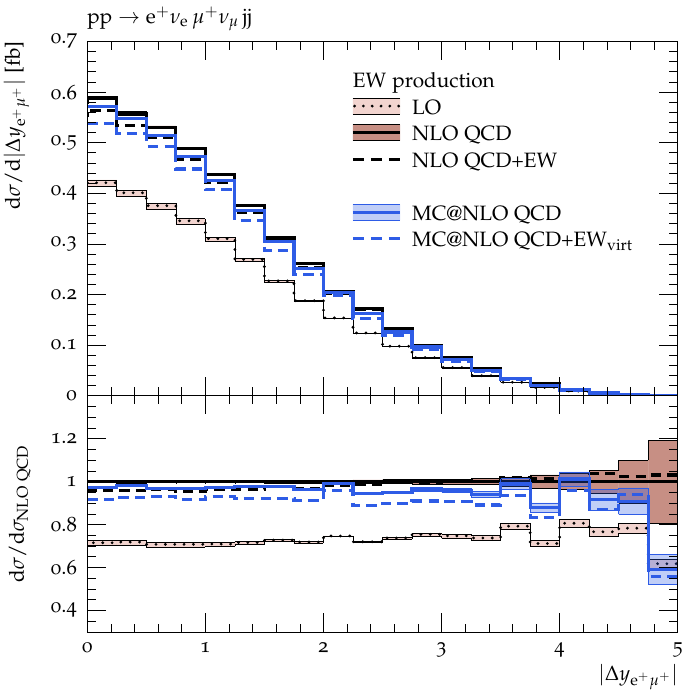}
  \end{subfigure}
  \\
  \begin{subfigure}{0.47\textwidth}
    \includegraphics[width=\textwidth]{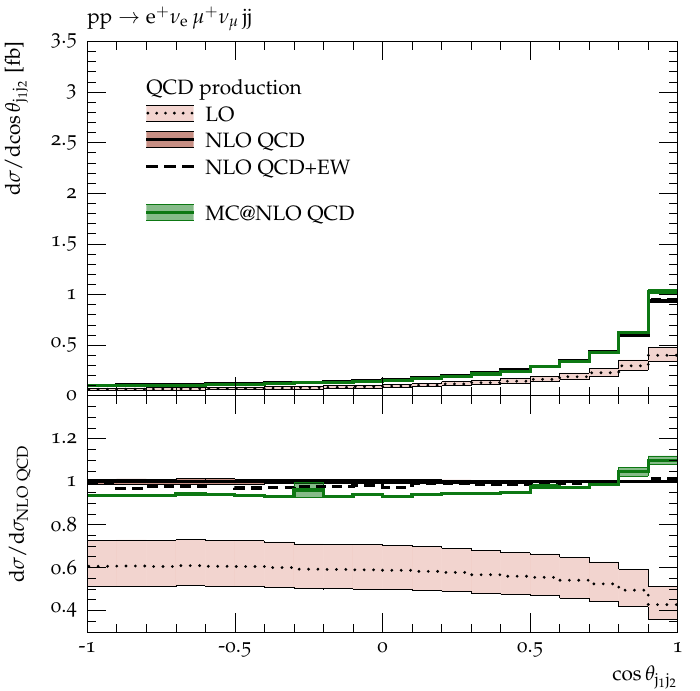}
  \end{subfigure}
  \begin{subfigure}{0.47\textwidth}
    \includegraphics[width=\textwidth]{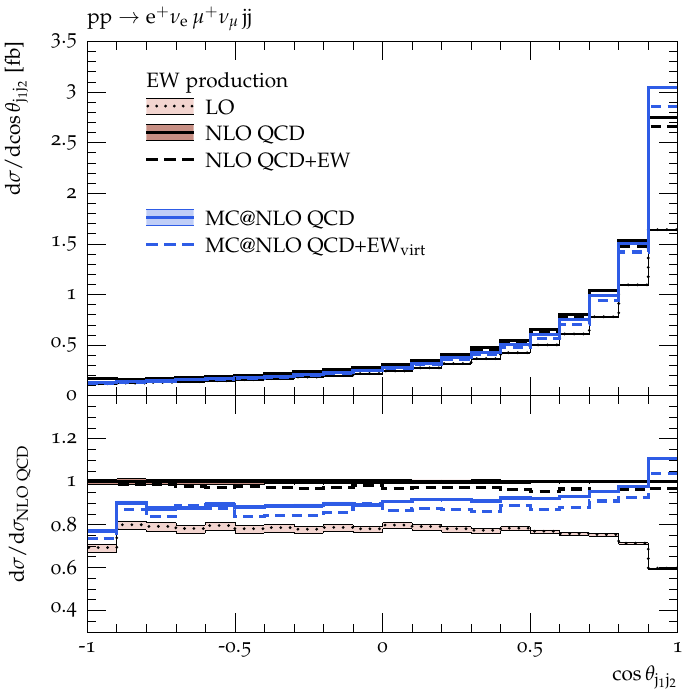}
  \end{subfigure}
  \caption{
    Parton-shower-matched predictions for the QCD and EW production
    modes contributing to $\process$ in comparison to LO and NLO QCD
    and NLO QCD+EW predictions. Results are shown for  the modulus of the
    rapidity difference between the charged leptons (top row) and
    the cosine of the angle between the two leading jets (bottom row). 
    \label{fig:mcatnlo-dy-ct}
  }
\end{figure}
We turn our discussion to differential distributions of physical
observables. To this end, \reffis{fig:mcatnlo-pT}--\ref{fig:mcatnlo-dy-ct}
contrast the fixed-order predictions with \MCatNLO simulations for both the
QCD and EW production processes in the left and right panels, respectively.
At fixed order we show corresponding results at LO (dotted), NLO QCD (solid)
and NLO QCD+EW (dashed). For the shower simulations we present results at
\MCatNLO accuracy, where for the EW production mode we furthermore include
EW corrections in the \EWvirt\ approximation. 

We begin the discussion of differential distributions with the transverse momentum of the
anti-muon in the  top-left and top-right panels of \reffi{fig:mcatnlo-pT}.
Here we observe that the shower-matched calculation predicts an
increase in the differential cross section of 10\% or
more for $p_{\mathrm{T},\mu^+}\gtrsim 250\,\text{GeV}$ in
the QCD production mode. This difference roots in the different arguments
used for the strong coupling in the fixed-order calculation and the \MCatNLO
simulation. For the former the unique scale $\muR$ is determined by Eq.~\refeq{eq:scale},
which for large $p_{\mathrm{T},\mu^+}$ gets large. In the matched calculation, however,
strong-coupling factors get evaluated at the respective jet-emission scales,
which are significantly smaller than $m_{\rT, \Pj\Pj} + m_{\rT, \nu_\Pe \Pe^+}
+ m_{\rT, \nu_\mu \mu^+}$, resulting in a relative enhancement of such events
in the \MCatNLO prediction. For the EW channel, on the other hand, both the 
fixed-order calculation and the shower-matched one agree
very well for $p_{\mathrm{T},\mu^+}\gtrsim 70\,\text{GeV}$ but
deviate by about 5\% for smaller $p_{\mathrm{T},\mu^+}$ values,
in agreement with the differences observed for the inclusive
cross section. Notably, the exact EW corrections are well-reproduced by the
\EWvirt\ approximation for $p_{\mathrm{T},\mu^+}\gtrsim 70\,\text{GeV}$.

In the lower-left and lower-right panels, \reffi{fig:mcatnlo-pT} furthermore
shows the transverse momentum of the leading-jet pair in the QCD and EW
production mode, respectively. The \MCatNLO calculation predicts a slightly larger
cross section at high-$p_{\mathrm{T,\Pj\Pj}}$ in the QCD channel, again related
to the above explained differences in the scale-setting prescription, while it
is generally well reproduced in the EW production process for
$p_{\rT,\Pj\Pj}\lesssim300\GeV$. 
However, the EW corrections now show a different behaviour.
For the QCD production process they are positive and accidentally
of a similar size as the QCD parton-shower corrections. For the EW production
they are negative, but in fact not well reproduced by the \EWvirt\ approximation.
The exact NLO EW corrections are only half of what their
Sudakov-approximation in the \EWvirt\ approximation suggests,
with the difference being made up by photon-induced contributions
missing in the \EWvirt\ ansatz.

Turning to invariant-mass distributions in \reffi{fig:mcatnlo-m},
the di-jet invariant mass displays again many aspects of the dynamics
of this process.
While being generally featureless and rather flat in the QCD production mode,
the fixed-order calculation agrees quite well with the shower-matched one.
For invariant masses above $\MW$ we observe a reduction at the level of $5\%$.
In contrast, for smaller values of $m_{\mathrm{\Pj_1\Pj_2}}$ we find a notable
increase in cross section reaching up to 15\%. In general, out-of-cone
final-state radiation will shift the di-jet invariant mass to somewhat lower
values, here leading to an accumulation of events where the rather flat
distribution starts to drop off. 
The EW corrections introduce a characteristic structure
through the interference of diagrams with $s$-channel resonances and
diagrams without them. These contributions do not give rise to a
Breit--Wigner peak. Instead, a dip--peak structure emerges in the
NLO QCD+EW prediction, \ie we observe a depletion of the cross section
right below the \PW mass and an increase just above it. While this structure integrates
to a very small inclusive contribution (for a vanishing \PW width this would
form an integrable singularity), its impact on the differential distribution
is rather sizeable. In the \MCatNLO QCD+\EWvirt\ algorithm, such mixed QCD--EW
correction would be applied prior to matching the QCD emission to the shower,
and would thus be smeared throughout the entire phase space. This rather unwanted
behaviour leads us to abandon the \EWvirt\ approximation for the \MCatNLO
description of the QCD production process.

The EW production process, on the other hand, features the full
Breit--Wigner shape at the \PW mass already at LO.
The NLO QCD corrections are very sizeable and basically overshadow
any potential feature from mixed QCD--EW contributions. The separation of the
\MCatNLO calculation in $s$- and $t/u$-channel processes removes this feature
as well, and the resulting shower-matched result agrees rather well with the
fixed-order one. Notable differences, however, induced by multiple-emission effects,
can be seen in the region below \MW, and the further distortion of the shape of
the distribution around the \PW-mass peak itself. As for the QCD process,
multiple shower emissions migrate events from higher to lower invariant
masses, resulting in about 15\% higher predicted cross section in the \MCatNLO
calculation below the Breit--Wigner peak, \ie for $m_{\Pj_1\Pj_2}$ between 20 and
60\,GeV. The EW corrections are moderate in this observable and well reproduced
at small invariant masses. For $m_{\Pj_1\Pj_2}\gtrsim \MW$, the exact EW corrections
are small but positive, while the \EWvirt\ approximation
predicts small negative corrections.
Again, the difference originates in the photon-induced contributions
only present in the exact calculation.
This is nevertheless unproblematic,
as the EW corrections are moderate and this
observable is outside the validity of the \EWvirt\ approximation throughout
its range considered here.

The bottom two plots of \reffi{fig:mcatnlo-m} show the invariant-mass
distribution of the system formed by the two leading jets and the
positron and anti-muon. Similarly to the $p_\mathrm{T}$-type distributions
in \reffi{fig:mcatnlo-pT},  for the QCD
production channel the difference between NLO and \MCatNLO predictions
is within $\sim5\%$  for small invariant masses, but increases to $20\%$
for higher values of $m_{\Pj_1\Pj_2\Pe^+\mu^+}$.
As before, the EW corrections are miniscule throughout the
investigated range.
For the EW production process, the agreement between the NLO and
\MCatNLO predictions is worse, driven entirely by the $t/u$-channel
contribution. For $m_{\Pj_1\Pj_2\Pe^+\mu^+}\gtrsim 500\,\text{GeV}$, the further
increase in the cross section through multiple emissions described
only in the shower-matched calculation modifies the spectrum at
the level of up to 50\%.
In the exact fixed-order calculation, the net EW corrections are
very small, exceeding $\pm 5\%$ only in the very first bin. 
This roots in the fact that their large Sudakov logarithms are calculated
as a correction to the Born process.
Their impact is then countered with moderately sized
positive photon-induced contributions, leading to a small overall effect.
On the contrary, the \MCatNLO matched \EWvirt\ approach effectively applies the
large Sudakov logarithms on showered events that also carry the bulk of
large QCD corrections.
As a consequence, the effect of the Sudakov corrections is amplified
by the additional QCD corrections of the shower.
Nonetheless, the \EWvirt\ approach still misses positive photon-induced
corrections of up to 8\% in the spectrum.

Finally, \reffi{fig:mcatnlo-dy-ct} shows the distributions in the rapidity separation of
the two charged leptons and the cosine of the opening angle between
the two jets for both production modes.
For the leptonic observable, both for the QCD and EW production mode
the NLO predictions agree well with the \MCatNLO calculations. In
particular, for the QCD channel the EW corrections are very
small. Also in the EW channel they are very moderate and well reproduced
by the \EWvirt\ approximation, so their overall size is larger by
about $6\%$, as already seen for the total cross sections in \refta{tab:MCNLO}.

The situation is somewhat different for the angle between the QCD
jets. Already for the QCD channel we observe an enhancement of
events with $\cos\theta_{\Pj_1\Pj_2}\approx 1$, \ie rather collinear
jets, at the expense of a slight suppression for larger angles,
\ie small $\cos\theta_{\Pj_1\Pj_2}$. This effect can be traced back to
collinear shower emissions off the hard-process partons that ultimately
form one of the two leading jets entering the distribution. The impact of
multiple shower emissions is even more pronounced for the EW production
mode. Here the suppression on the left tail of the distribution
reaches up to 10\% relative to the NLO QCD prediction, while the bin
close to $\cos\theta_{\Pj_1\Pj_2}=1$ gets an enhancement of the same
size. On the other
hand, the EW corrections both for the fixed order and the \MCatNLO
calculation are rather constant and uniform throughout almost the
entire observable range. Their size being again somewhat larger for
the \MCatNLO prediction, in line with the results shown in
\refta{tab:MCNLO}.

\section{Conclusion}
\label{sec:conclusion}

This article presents a detailed study of the process $\process$ at the LHC.
While this final state is usually associated to vector-boson
scattering and its same-sign channel, it also involves $\PW\PW\PW$ production which has
been analysed in specific experimental measurements. The tri-boson
contributions are, in particular, relevant for the study of anomalous
quartic gauge-boson couplings.

In this article, we have first studied the various production
mechanisms at LO and NLO accuracy and find that for the considered
phase space, targeted to tri-boson measurements, a large fraction (about $40\%$) of the cross section can actually be attributed to $\PW\PH$ production.
Obviously, the $\mu^+\nu_\mu\Pe^+\nu_{\Pe}\Pj\Pj$ final state is far from trivial as it contains many intricate production mechanisms.

We have further computed the full NLO corrections to the off-shell production.
This allows us to get a deep understanding of the structure of higher-order corrections for this final state.
In particular, we reconciliate a-priori contradicting observations:
tri-boson production has relatively small EW
corrections, while vector-boson scattering has intrinsically large ones.
We confirm that both statements are correct but point out that
different phase-space regions enhance production mechanisms that have different dynamics and therefore different EW corrections.
The higher-order corrections can reach $50\%$ at the level of the total cross section and show a very different hierarchy with respect to the one in VBS phase spaces.
At the level of differential distributions, the corrections exceed
$100\%$ in certain phase-space regions owing to real radiation.

Differences between the off-shell and on-shell calculations are at the
level of few per cent for fiducial cross sections. In distributions,
deviations are found at the level of $10\%$ for large transverse
momenta and up to $60\%$ in invariant-mass distributions away from the
resonances.

Moreover, we provide NLO-QCD matched predictions for the EW and QCD
production mode supplemented  with approximate EW corrections for the
former, where they are relevant. Inclusion of the parton shower reduces the fiducial cross
section by $6$--$8\%$. Differential distributions are modified at the
level $10$--$20\%$ in phase-space regions with appreciable cross
sections. The \EWvirt\ approximation performs reasonably well in
capturing the dominant NLO EW corrections for the considered phase space.
To further improve the shower-matched simulations, photon-induced
contributions would need to be included, as these can yield sizeable
effects in the tails of some observable distributions.

All in all, the present study delivers state-of-the-art predictions
for $\process$ at the LHC. We hope that they will foster 
comparisons between Standard Model predictions and experimental data in the future.
It is worth stressing that all matched predictions have been obtained with the general-purpose Monte Carlo {\sc Sherpa} and that they can readily be reproduced and used in experimental analyses.

Finally, we would like to point out that the present work illustrates
perfectly the richness of the interplay between experimental data and theoretical predictions. In particular, the interpretation of the experimental data in terms of simple production mechanisms turned out to be non-trivial and therefore requires an intricate work between the experimental and theory community.

\section*{Acknowledgements}

\begin{sloppypar}
AD acknowledges financial support by the German
Federal Ministry for Education and Research (BMBF) under contract
no.~05H21WWCAA.
MP acknowledges support by the German Research Foundation (DFG) through the Research Training Group RTG2044 and through grant no.~INST 39/963-1 FUGG (bwForCluster NEMO) as well as the state of Baden-Württemberg through bwHPC.
MS is funded by the Royal Society through a University Research Fellowship
(URF\textbackslash{}R1\textbackslash{}180549, URF\textbackslash{}R\textbackslash{}231031) and an Enhancement Award
(RGF\textbackslash{}EA\textbackslash{}181033,
 CEC19\textbackslash{}100349, and RF\textbackslash{}ERE\textbackslash{}210397)
as well as the STFC (ST/X003167/1 and ST/X000745/1).
The work of SS was supported by BMBF (contract no.~05H21MGCAB) and
DFG (project no.~456104544).
\end{sloppypar}

\bibliographystyle{utphys.bst}
\bibliography{triboson}
\end{document}